\documentclass[12pt]{article}
\usepackage{amsmath}
\usepackage{graphicx}
\usepackage{enumerate}
\usepackage{natbib}
\usepackage{url} 
\usepackage[section]{placeins}

\usepackage{amsfonts}
\usepackage{amsthm}
\usepackage{multirow}   
\usepackage{setspace}
\usepackage{amssymb}
\usepackage{amsmath}
\usepackage{graphicx,color,epstopdf}
\usepackage{hyperref}
\usepackage{xcolor}
\usepackage{booktabs}
\usepackage{subfigure}
\usepackage{array}

\usepackage{dsfont }  
\usepackage{algorithm}  
\usepackage{algorithmic}
\usepackage{enumerate}     
\usepackage{multirow}
\usepackage{setspace}

\usepackage{float}

\usepackage{graphicx}

\newtheorem{Th}{\bf Theorem}
\newtheorem{remark}{Remark}
\newtheorem{con}{Condition}
\newcommand{\blind}{1}

\numberwithin{equation}{section}

\newcommand{\bSigma}{\boldsymbol{\Sigma}}
\newcommand{\bbeta}{\boldsymbol{\beta}}

\newcommand{\btheta}{\boldsymbol{\theta}}
\newcommand{\bTheta}{\boldsymbol{\Theta}}

\begin{document}

	\def\spacingset#1{\renewcommand{\baselinestretch}%
		{#1}\small\normalsize} \spacingset{1}

	
	\if1\blind
	{
		\title{\bf The conditionally studentized test for
			high-dimensional parametric regressions
		}
		\author{Feng Liang$^{1}$, Chuhan Wang$^{1}$, Jiaqi Huang$^{2}$ and Lixing Zhu$^{1,3}$
			\footnote{All the authors contribute to this research equally and the names are in seniority order. Jiaqi Huang is the corresponding author. Corresponding author (J. Huang). Email addresses: jhuang@mail.bnu.edu.cn (J. Huang).}\\
			$^1$ Center for Statistics and Data Science, Beijing Normal University, Zhuhai,
			China\\
			$^2$ School of Statistics, Beijing Normal University, Beijing, China\\
			$^3$ Department of Mathematics, Hong Kong Baptist University, Hong Kong\\
		}
		\maketitle
	} \fi
	
	\if0\blind
	{
		\bigskip
		\bigskip
		\bigskip
		\begin{center}
			{\LARGE\bf The conditionally studentized test for
			high-dimensional parametric regressions}
		\end{center}
		\medskip
	} \fi
	
	\bigskip
	\begin{abstract}
		This paper studies model checking for general parametric regression models having no dimension reduction structures on the predictor vector. Using any U-statistic type test as an initial test, this paper combines the sample-splitting and conditional studentization approaches to construct a COnditionally Studentized Test (COST). Whether the initial test is global or local smoothing-based; the dimension of the predictor vector and the number of parameters are fixed or diverge at a certain rate, the proposed test always has a normal weak limit under the null hypothesis. When the dimension of the predictor vector diverges to infinity at faster rate than the number of parameters, even the sample size, these results are still available under some conditions. This shows the potential of our method to handle higher dimensional problems. Further, the test can detect the local alternatives distinct from the null hypothesis at the fastest possible rate of convergence in hypothesis testing. We also discuss the optimal sample splitting in power performance. The numerical studies offer information on its merits and limitations in finite sample cases including the setting where the dimension of predictor vector equals the sample size. As a generic methodology, it could be applied to other testing problems.
	\end{abstract}
	
	\noindent%
	{\it Keywords:}  Asymptotic model-free test; conditional studentization; high dimensions; model checking; sample-splitting.
	\vfill
	
	\newpage
	\spacingset{1.9} 
	\section{Introduction}\label{sec1}
	Consider the general  parametric regression model:
	$$ Y=g(\boldsymbol{X};\btheta)+\varepsilon,$$ where $g$ is a known function of $ (\boldsymbol{X}, \btheta)$ and $\btheta \in \bTheta \subset \mathbb{R}^{p}  $ is an unknown parameter vector. $\boldsymbol{X}$ is a predictor vector in $\mathbb{R}^{q}$ while $Y$ represents the univariate response variable where $\mathbb{R}^{p}$ ($\mathbb{R}^{q}$) stands for the $p$-($q$-)dimensional Euclidean space. For many models, such as linear and generalized linear models, $p=q$, but in general, this may not be the case. As well known, when the assumed model is used, model checking is necessary before further analyzing data.  Specifically, the null hypothesis is, for a subset $\bTheta \subset \mathbb{R}^{p}$,
	\begin{eqnarray}\label{0101}
		\label{h0}
		H_0: \Pr\{E(Y|\boldsymbol{X})=g(\btheta_0,\boldsymbol{X})\}=1 \text{\ for some\ } \btheta_0 \in \bTheta
	\end{eqnarray}
	versus the alternative hypothesis
	\begin{eqnarray}
		\label{h1}
		H_1: \Pr\{E(Y|\boldsymbol{X})=g(\boldsymbol{\theta},\boldsymbol{X})\}<1 \text{\ for all\ } \btheta \in \bTheta.
	\end{eqnarray}
	Relevant problems have been investigated intensively in the literature. Most existing methods are for cases with fixed dimensions $p$ and $q$. Examples include 
	nonparametric estimation-based local smoothing tests such as  \cite{hardle1993comparing}, \cite{zheng1996consistent}, \cite{zhu1998dimension}, \cite{lavergne2008breaking,lavergne2012one}, \cite{guo2016model}, and empirical process-based global smoothing tests,  e.g. \cite{stute2002model}, \cite{zhu2003model}, \cite{escanciano2006consistent}, and \cite{stute2008model}.  These two general methodologies respectively show their advantages and limitations. Local smoothing tests often have tractable weak limits under the null hypothesis and most of them can detect local alternatives distinct from the null hypothesis at slower rates of convergence than $1/\sqrt n$ where $n$ is the sample size. As the rates can be slower in higher dimensional cases, the curse of dimensionality is a big challenge. In contrast, the limiting null distributions of global smoothing tests under the null hypothesis are usually  intractable, resorting to resampling approximations to determine critical values, but can detect local alternatives distinct from the null hypothesis at the fastest possible rate $1/\sqrt n$ of convergence in hypothesis testing. The dimensionality is still an issue as this type of test involves high dimensional empirical process.
	
	For  paradigms with large dimensions $p$ and $q$ that may diverge as the sample size goes to infinity,
	there are few methods available in the literature. Some relevant references for models with dimension reduction structures are cited as follows. \cite{shah2018goodness} and \cite{jankova2020goodness} respectively proposed two goodness-of-fit tests for high-dimensional sparse linear and generalized linear models with a fixed designs.   For the problems with random designs without sparse structures,  \cite{tan2019adaptive} and \cite{tan2022integrated} considered adaptive-to-model tests for high-dimensional single-index and multi-index models that the number of linear combinations of $\boldsymbol{X}$ is fixed, respectively, with diverging dimensions $p\geq q$. Their methods are the extensions of the test first proposed by \cite{guo2016model}  in fixed-dimension cases for multi-index models. These three tests critically hinge on the dimension reduction structures of the predictors. Otherwise, as \cite{tan2022integrated} discussed, they fail to work because the limiting distributions under the null and alternative hypothesis are degenerate at constants and cannot be used to determine critical values and resampling approximation cannot work well either. 
	
	The current paper proposes a  COnditionally Studentized Test (COST) for  general parametric models without dimension reduction structures in high-dimensional cases.  The basic idea for constructing this novel test is that based on an initial test that can be rewritten as a U-statistic (either a local or a global smoothing test), we divide the sample of size $n$ into two subsamples of sizes $n_1$ and $n_2$, and use the conditional studentization approach to construct the final test. The dimension $p$ can diverge at the rate with a leading term $n_1^{1/3}$ corresponding to the rate $n^{1/3}$ \cite{tan2022integrated}  achieved.  Further, the restriction that $q\le p$ is no longer necessary. That is, the dimension $q$ of the predictor vector can be higher than $p$, even the sample size $n$  under some  regularity conditions on the regression and related functions. Note that the conditions in this paper are not assumed on predictor significance to the regression function.  Thus,  understandably, when $q$ is large, the conditions could  relatively stringently restrict the forms of related functions, but the results still show the potential of our method  in  higher dimension settings.     The details will be presented in Section~\ref{sec2}. Remark~1 in this section also gives some more explanations. Section~\ref{sec4} reports some numerical studies to check the performance of the test when $q$ is larger than $p$ and equal to $n$.  As the conclusions are similar, some other studies with $q>n$ are not reported for space saving. The conditions are put in Section~\ref{sec6}. We also discuss the optimal sample splitting between the sizes $n_1$ and $n_2$ in the power performance of the test. The following merits of the novel test are worthy of mentioning. Under more general parametric model structures, whether the initial test is global or local smoothing-based, and whether the dimensions are fixed or divergent at certain rates, the final test can always have a normal weak limit under the null hypothesis, which is often a merit of local smoothing tests;  and can detect local alternatives distinct from the null at the rate as close to $1/\sqrt {n_1}$ (${n_1}/n\to C $ for a given positive constant $C$) as possible, which is the typical optimal rate global smoothing tests can achieve.  These unique features are very different  from any existing test as, other than being able to handle high-dimensional problems, the test also enjoys the advantages local smoothing tests and global smoothing tests have.    On the other hand,  the test statistic converges to its weak limit at the rate of $1/\sqrt {n_1}$ rather than $1/\sqrt {n}$ such that it may lose some power in theory.   We will discuss    this limitation  in more detail later.

	The rest of the paper is organized as follows.  Section 2 describes  the test statistic construction. Section \ref{sec3} includes the asymptotic properties of the test statistic under the null and alternative hypotheses, the investigation on  the optimal choice of sample-splitting scheme. Section \ref{sec4} contains some numerical studies  including simulations and real data analysis. To examine its performance,  the simulation studies include  the settings  favoring the existing method in the literature; the settings where the condition on the number of parameters is violated; and the settings with the dimension of predictor vector being much larger than the number of parameters, even equal to the sample size.   Section \ref{sec5} comments on  its advantages and limitations. It gives a discussion of the reason why in our setting,  we do not apply the commonly used cross-fitting approach for power enhancement, and a brief discussion on the challenge of extending or modifying our method to handle  models having higher dimensions with sparse structure. Section~\ref{sec6} includes the regularity conditions with some remarks. As the proofs of the main results are technically demanded and lengthy,  we then put them in Supplementary Material.

	\section{Test statistic construction}\label{sec2}
	As the test construction requires estimating the parameter in the model, we first briefly give the details.
	\subsection{Notation and parameter estimation}

	Write the underlying regression function as $m(\boldsymbol{x})=E(Y|\boldsymbol{X}=\boldsymbol{x})$ and the error   as  $\varepsilon=Y-m(\boldsymbol{X})$.
	To study the power performance, we also consider a sequence of alternative hypotheses:
	\begin{equation}\label{alter}
		H_{1 n}: Y=g(\btheta_0,\boldsymbol{X})+\delta_{n} l(\boldsymbol{X})+\varepsilon.
	\end{equation}
	When $\delta_n\to 0$ as $n\to\infty$, (\ref{alter}) corresponds to a sequence of local alternatives. When $\delta_n$ is a fixed constant, (\ref{alter}) reduces to the global alternative model (\ref{h1}).
	
	Set
	\begin{eqnarray}\label{theta*}
		\btheta^*=\underset{\btheta \in \boldsymbol{\Theta}}{\arg \min } E\left\{Y-g(\btheta, \boldsymbol{X})\right\}^{2}=\underset{\btheta \in \bTheta}{\arg \min } E\left\{m(\boldsymbol{X})-g(\btheta, \boldsymbol{X})\right\}^{2}.
	\end{eqnarray}
	Under the null hypothesis in (\ref{0101}) and the regularity condition \ref{CA.1} specified in Section \ref{sec6}, $\btheta^*=\btheta_{0}$. Under the alternatives, $\btheta^*$ typically depends on the  distribution of $\boldsymbol{X}$. To save space, redefine $l(\boldsymbol{X})=m(\boldsymbol{X})-g(\boldsymbol{\theta}^*,\boldsymbol{X})$ under the global alternative hypothesis with fixed $\delta_n=1$, while we still write $l(\boldsymbol{X})=m(\boldsymbol{X})-g(\boldsymbol{\theta}_0,\boldsymbol{X})$ under the local alternative hypothesis.

	Subsequently, we proceed to present additional notations that have been employed in this paper. Denote $\dot{g}(\btheta, \boldsymbol{X})=\partial g(\btheta, \boldsymbol{X})/\partial \btheta$ and $\ddot{g}(\btheta, \boldsymbol{X})=\partial \dot{g}(\btheta, \boldsymbol{X})/\partial \btheta^{\top}$.
	Under the null hypothesis and the local alternative hypothesis,  we define $\bSigma =E\left\{\dot{g}(\btheta_0, \boldsymbol{X}) \dot{g}(\btheta_0, \boldsymbol{X})^{\top}\right\}$. Under the global alternatives, without confusion, we define $\bSigma =E\left\{\dot{g}(\btheta^*, \boldsymbol{X}) \dot{g}(\btheta^*, \boldsymbol{X})^{\top}\right\}-E\left[\left\{m(\boldsymbol{X})-g(\btheta^*,\boldsymbol{X})\right\}\ddot{g}(\btheta^*, \boldsymbol{X})\right]$ and $\bSigma_*=E\left\{\dot{g}(\btheta^*, \boldsymbol{X}) \dot{g}(\btheta^*, \boldsymbol{X})^{\top}\right\}$.
	Use $\|\cdot\|$ to represent the $L_2$ norm. Write the conditional expectation $E(A|B=b)$ as $E_B(A)$ for any random variable $A$ and random variable/vector $B$.
	
	The least squares estimator of $\btheta$ is defined by
	\begin{eqnarray}\label{thetah}
		\boldsymbol{\hat{\theta}}=\underset{\btheta \in \bTheta}{\arg \min } \sum_{i=1}^{n}\left\{Y_{i}-g(\btheta, \boldsymbol{X}_{i})\right\}^{2}.
	\end{eqnarray}
	Under the regularity conditions \ref{CA.1}-\ref{CA.4} in Section \ref{sec6},   the convergence rate and  asymptotically linear representation of  $\boldsymbol{\hat{\theta}}$ can be derived, which are important for studying the asymptotic properties of the test statistic under the null and alternative hypotheses. We will state them as three lemmas in Supplementary Material, and the first two lemmas are Theorems 1 and 2 and the third lemma is an extension of Theorem~4  in  \cite{tan2022integrated}.
	
	\subsection{The motivation}
	In the literature,  several existing tests have a similar structure, with a weight function  $W_n(\cdot, \cdot)$ depending on the sample size $n$ such that a test statistic can be written as, before standardizing, $ \sum_{i=1}^{n} \sum_{j \neq i} \hat{e}_{i} \hat{e}_{j} W_n(\boldsymbol{X}_i, \boldsymbol{X}_j)/n(n-1)$ where  $\hat{e}_{i}=Y_i-  g( \boldsymbol{\hat{\theta}}, \boldsymbol{X}_i)$ is the residual. Some classic tests are listed as follows. \cite{bierens1982consistent} proposed the integrated conditional moment (ICM) test based on the weight function $\exp(-\|\boldsymbol{X}_1-\boldsymbol{X}_2\|^2/2)$;   The tests proposed by \cite{escanciano2009lack} discussed the case with general weight functions; \cite{li2019model}  used the weight function $1/\sqrt {\|\boldsymbol{X}_1-\boldsymbol{X}_2\|^2+1}$ induced by an idea bridging local and global smoothing tests. Other tests can be written or approximately written as U-statistics including those local smoothing tests proposed by \cite{hardle1993comparing}, \cite{zheng1996consistent}, and those global smoothing tests suggested by  \cite{stute1998model,stute1998bootstrap},  \cite{tan2019adaptive} and \cite{tan2022integrated}. However,  these tests  do not apply to the cases with diverging dimensions $p$ without dimension reduction structures.
	
	We now construct a novel test by combining the sample-splitting and the conditional studentization approaches.  The following observations induce the test construction. Let $e=Y-g(\boldsymbol{\theta}_0,\boldsymbol{X})$.   Under the null hypothesis,  $m(\boldsymbol{X})=g(\boldsymbol{\theta}_0,\boldsymbol{X})$ and $e=\varepsilon$.   Note that  $E(e|\boldsymbol{X})=0$ under the null hypothesis, that is, $E\left\{e E(e|\boldsymbol{X})f(\boldsymbol{X})\right\}=E\left\{E^2(e|\boldsymbol{X})f(\boldsymbol{X})\right\}=0$, otherwise greater than zero under the alternatives. In a more general presentation,  when we use an approximation $E_{\boldsymbol{X}}\{e W_n(\boldsymbol{X}, \boldsymbol{X}_2)\}$ with, say, a kernel function $K\left\{(\boldsymbol{X}-\boldsymbol{X}_2)/h\right\}/h^{q}$ in lieu of $W_n(\boldsymbol{X}, \boldsymbol{X}_2)$   to $E(e|\boldsymbol{X})f(\boldsymbol{X})$,  this quantity can be approximated by $E\left\{e_1 e_2 W_n(\boldsymbol{X}_1, \boldsymbol{X}_2)\right\}$. Simply, we abbreviate $W_n(\boldsymbol{X}_i,\boldsymbol{X}_j)$ as $w_{ij}$. Thus, in general, this quantity can be estimated by  $\frac{1}{n}\sum_{i=1}^{n}\hat{e}_{i}\left(\frac{1}{n-1}\sum_{j=1, j\not = i}^{n}\hat{e}_{j}w_{ij}\right)=\frac{1}{n(n-1)}\sum_{i=1}^{n}\sum_{j=1, j\not = i}^{n}\hat{e}_{i}\hat{e}_{j}w_{ij} $ that can be used to define a non-standardized statistic:
	\begin{align}
		\label{U1}
		U_n=\frac{1}{\sqrt{n}}\sum_{i=1}^{n}\hat{e}_{i}\left(\frac{1}{\sqrt{n-1}}\sum_{j=1, j\not = i}^{n}\hat{e}_{j}w_{ij}\right)=\frac{1}{\sqrt{n(n-1)}}\sum_{i=1}^{n}\sum_{j=1, j\not = i}^{n}\hat{e}_{i}\hat{e}_{j}w_{ij}.
	\end{align}
	Different tests uses different weight functions $w_{ij}$. Examples include the
	kernel function $K\left\{(\boldsymbol{X}_i-\boldsymbol{X}_j)/h\right\}/h^{q}$ with a bandwidth $h$ (see, e.g., \cite{zheng1996consistent}),  exponential function  $\exp(-\|\boldsymbol{X}_i-\boldsymbol{X}_j\|^2/2)$ (see e.g., \cite{bierens1982consistent}), and the weight function $1/\sqrt {\|\boldsymbol{X}_i-\boldsymbol{X}_j\|^2+1}$  used by \cite{li2019model}. 
	These tests often have no tractable limiting null distributions.
	
	\subsection{The test statistic}
	We now modify   $U_n$ to avoid  duplicated use of the samples. Estimate $\btheta$ by two parts of samples respectively. Define
	$$\boldsymbol{\hat{\theta}}_1=\underset{\btheta \in \bTheta}{\arg \min } \sum_{i=1}^{n_1}\left\{Y_{i}-g(\btheta, \boldsymbol{X}_{i})\right\}^{2} \text{ and } \boldsymbol{\hat{\theta}}_2=\underset{\btheta \in \bTheta}{\arg \min } \sum_{j=n_1+1}^{n}\left\{Y_{j}-g(\btheta, \boldsymbol{X}_{j})\right\}^{2},$$
	where the samples of size $n$ are divided into two disjoint parts $ \mathcal{N}_{1}=\{(\boldsymbol{X}_i,Y_{i})\}_{i=1}^{n_{1}} \text { and } \mathcal{N}_{2}=\{(\boldsymbol{X}_j, Y_{j})\}_{j=n_{1}+1}^{n}$ of sizes $n_1$  and $n_2$ satisfying $n=n_1+n_2$.  All the results in Section~\ref{sec2} hold when $n$ is replaced by $n_1$ or $n_2$. Define a modified test statistic as
	\begin{equation}
		\label{U2} U_{\mathcal{N}_{1},\mathcal{N}_{2}}=\frac{1}{\sqrt{n_1}}\sum_{i=1}^{n_1}\hat{e}_{i}\left(\frac{1}{\sqrt{n_2}} \sum_{j=n_1+1}^{n}\hat{e}_{j}w_{ij}\right)=\frac{1}{\sqrt{n_1n_2}}\sum_{i=1}^{n_1}\hat{e}_{i} \sum_{j=n_1+1}^{n}\hat{e}_{j}w_{ij},
	\end{equation}
	where $\hat{e}_{i}=Y_i-g(\boldsymbol{\hat{\theta}}_1,\boldsymbol{X}_i), \ i=1, \dots,n_1$ and $\hat{e}_{j}=Y_j-g(\boldsymbol{\hat{\theta}}_2,\boldsymbol{X}_j), \ j=n_1+1, \dots, n.$
	Again, this test statistic usually has  no significant difference from the previous $U_n$ as they have similar asymptotic behaviors. However, this seemingly minor modification plays a vital role in constructing a conditionally studentized test with a normal weak limit under the null hypothesis. The key ingredient in the construction is that in the two independent sums the residuals are not used duplicately, but the weight function $w_{ij}$ links the two sums.
	
	
	To see how to define the final statistic, we give the decomposition of $U_{\mathcal{N}_{1},\mathcal{N}_{2}}$. Under the null hypothesis, we have the followings:
	\begin{align}
		U_{\mathcal{N}_{1},\mathcal{N}_{2}}&=\frac{1}{\sqrt{n_1n_2}}\sum_{i=1}^{n_1}\sum_{j=n_1+1}^{n}\hat{e}_{i}\hat{e}_{j} w_{ij}\nonumber\\
		&=\frac{1}{\sqrt{n_1}}\sum_{i=1}^{n_1}e_i\frac{1}{\sqrt{n_{2}}}\sum_{j=n_{1}+1}^{n} \hat{e}_{j}\left[w_{ij}-\dot{g}(\btheta^*, \boldsymbol{X}_{i})^{\top} \bSigma^{-1}E_{\boldsymbol{X}_j}\left\{\dot{g}(\btheta^*, \boldsymbol{X}_1) w_{1j}\right\}\right]+o_p(1)\nonumber\\
		&=:\frac{1}{\sqrt{n_1}}\sum_{i=1}^{n_1}e_i\tilde{w}_i(\mathcal{N}_2)+o_p(1),
	\end{align}
	where $E_{\boldsymbol{X}_j}\left\{\dot{g}(\btheta^*, \boldsymbol{X}_1) w_{1j}\right\}$ is the conditional expectation given $\boldsymbol{X}_j$, and $\btheta^*$ is defined in (\ref{theta*}). The detailed justification can be found in \textbf{Corollary 1} of Supplementary Material.
	The decomposition is only about the residuals $\hat e_i$'s in the first  part of the samples.  Given $\mathcal{N}_{2}$, $e_1\tilde{w}_1(\mathcal{N}_2), \ldots, e_{n_1}\tilde{w}_{n_1}(\mathcal{N}_2)$ are conditionally independent and identically distributed random variables. Intuitively, under the null hypothesis, the following random sequence would have a normal weak limit by using the Central Limit Theorem conditionally:
	\begin{equation}
		V_n:=\frac{\frac{1}{\sqrt{n_1}}\sum_{i=1}^{n_1}e_i\tilde{w}_i(\mathcal{N}_2)}{\sqrt{\frac{1}{n_1}\sum_{i=1}^{n_1}\left\{e_i\tilde{w}_i(\mathcal{N}_2)-\frac{1}{n_1}\sum_{i=1}^{n_1}e_i\tilde{w}_i(\mathcal{N}_2)\right\}^2}}
	\end{equation}
	that is a conditionally  studentized version of  $\sum_{i=1}^{n_1}e_i\tilde{w}_i(\mathcal{N}_2)$.
	To define the final test statistic,  we use
	$\sum_{i=1}^{n_1}\sum_{j=n_1+1}^{n}\hat{e}_{i}\hat{e}_jw_{ij}/\sqrt{n_1n_2}$ in lieu of $\sum_{i=1}^{n_1}e_i\tilde{w}_i(\mathcal{N}_2)/\sqrt{n_1}$ and for the denominator that is a conditional standard deviation, we use $\hat{e}_i$, $\hat{\bSigma}$ and $\boldsymbol{\hat{\theta}}$  to replace $e_i$, $\bSigma$ and $\btheta^*$:
	\begin{equation}
		\hat{V}_n:=\frac{\frac{1}{\sqrt{n_1n_2}} \sum_{i=1}^{n_1}\sum_{j=n_1+1}^{n}\hat{e}_{i}\hat{e}_jw_{ij}}{\sqrt{\frac{1}{n_{1}} \sum_{i=1}^{n_1}\left\{\hat{e}_{i} \tilde{\tilde{w}}_i(\mathcal{N}_2)-\frac{1}{n_{1}}\sum_{i=1}^{n_{1}}\hat{e}_{i} \tilde{\tilde{w}}_i(\mathcal{N}_2)\right\}^{2}}},
	\end{equation}
	where $$\tilde{\tilde{w}}_i(\mathcal{N}_2)=\frac{1}{\sqrt{n_2}}\sum_{j=n_{1}+1}^{n} \hat{e}_{j}\left[w_{ij}-\dot{g}\left(\boldsymbol{\hat{\theta}}, \boldsymbol{X}_{i}\right)^{\top} \hat{\bSigma}^{-1}\frac{1}{n_1}\sum_{i=1}^{n_1}\left\{\dot{g}\left(\boldsymbol{\hat{\theta}}, \boldsymbol{X}_i\right) w_{ij}\right\}\right],$$ and $\hat{\bSigma}=\frac{1}{n}\sum_{i=1}^{n}\dot{g}(\boldsymbol{\hat{\theta}},\boldsymbol{X}_i)\dot{g}(\boldsymbol{\hat{\theta}},\boldsymbol{X}_i)^{\top}$. We will prove that $\boldsymbol{\hat{\theta}}$, $\frac{1}{n_1}\sum_{i=1}^{n_1}\{\dot{g}(\boldsymbol{\hat{\theta}}, \boldsymbol{X}_i) w_{ij}\}$ and  $\hat{\bSigma}$ are the consistent estimators of $\btheta^*$, $E_{\boldsymbol{X}_j}\left\{\dot{g}(\btheta^*, \boldsymbol{X}_1) w_{1j}\right\}$ and $\bSigma_*$, respectively such that the consistency of this estimated  conditional standard deviation holds. Note that in $\tilde{\tilde{w}}_i(\mathcal{N}_2)$, we use the full data-based estimator $\boldsymbol{\hat{\theta}}$ instead of $\boldsymbol{\hat{\theta}}_2$. We find that asymptotically, there is no difference by using either estimator, but using $\boldsymbol{\hat{\theta}}$ can make a faster convergence rate of the estimator to $\boldsymbol{\theta}^*$. In \textbf{Corollary 2} of Supplementary Material, we will show that under the null hypothesis and the local alternative hypothesis in (\ref{alter}) with $np^3\delta_n^4\to 0$, 
	\begin{equation}
		\frac{\frac{1}{n_1}\sum_{i=1}^{n_1}\left\{e_i\tilde{w}_i(\mathcal{N}_2)-\frac{1}{n_1}\sum_{i=1}^{n_1}e_i\tilde{w}_i(\mathcal{N}_2)\right\}^2}{\frac{1}{n_{1}} \sum_{i=1}^{n_1}\left\{\hat{e}_{i} \tilde{\tilde{w}}_i(\mathcal{N}_2)-\frac{1}{n_{1}}\sum_{i=1}^{n_{1}}\hat{e}_{i} \tilde{\tilde{w}}_i(\mathcal{N}_2)\right\}^{2}}\stackrel{p}{\to}1,
	\end{equation}
	where $``\stackrel{p}{\to}"$ stands for convergence in probability.
	While, under the global alternative hypothesis in (\ref{alter}) with fixed $\delta_n$,  $\tilde{\tilde{w}}_i(\mathcal{N}_2)$  will be proved to be a consistent estimator of $\tilde{w}_i^{(0)}(\mathcal{N}_2):=\sum_{j=n_{1}+1}^{n} \hat{e}_{j}\left[w_{ij}-\dot{g}(\btheta^*, \boldsymbol{X}_{i})^{\top} \bSigma_*^{-1}E_{\boldsymbol{X}_j}\left\{\dot{g}(\btheta^*, \boldsymbol{X}_1) w_{1j}\right\}\right]/\sqrt{n_2}$.  
	Note that when a kernel function $K\left\lbrace (\boldsymbol{X}-\boldsymbol{X}_2)/h\right\rbrace /h^{q}$ is used as the weight function $W_n(\boldsymbol{X}, \boldsymbol{X}_2)$,  it can go to infinity as the sample size goes to infinity. But this is not a problem in our construction as the studentized test is scale-invariant, and the quantity $h^{q}$ in the weight $w_{ij}$  is eliminated from the numerator and denominator. Thus, we can consider the weight function without this quantity.
	
	\begin{remark}
		It is worth noticing that  in the theorems and corollaries, we  impose some restrictions on the divergence rate of the parameter dimension $p$ such as $p^3\log n_1/n_1\to 0$ or $np^3\delta_n^4\to 0$, but  do not directly give the constraints on the  dimension $q$ of the predictor vector $\boldsymbol{X}$.  In fact, some constraints are hidden in the regularity conditions in Section \ref{sec6} on the regression and other related   functions such that $p\geq q$ in \cite{tan2022integrated} is not required and  $q$ can diverge to infinity much faster than $p$, even the sample size $n$. For instance, our test statistic can deal with the following model:
		\begin{align*}
			Y= & \theta_1 X_1 X_2 \cdots X_{p^3} X_{p^3+1}+\theta_2 X_2 X_3 \cdots X_{p^3+1} X_{p^3+2} \\
			& +\theta_3 X_3 X_4 \cdots X_{p^3+2} X_{p^3+3}+\cdots+\theta_p X_p X_{p+1} \cdots X_{p^4-1} X_{p^4}+\varepsilon,
		\end{align*}
		where $\boldsymbol{X}=\left(X_1, X_2, \cdots, X_{p^4}\right) \sim \mathbf{N}\left(0, \mathrm{I}_{p^4}\right)$ and $q=p^4=n$. The required conditions are satisfied. Similarly, we can see some models with higher dimension $p> n$. Therefore, although the conditions are strong, the results show the potential of our method to handle the problems in large-dimensional settings.
	\end{remark}

	\section{Asymptotic properties}\label{sec3}
	\subsection{The limiting null distribution}
	Under the null hypothesis, the conditionally studentized version $V_n$ of $\sum_{i=1}^{n_1}e_i\tilde{w}_i(\mathcal{N}_2)$ has a normal weak limit under some regularity conditions (see  \textbf{Corollary 3} in Supplementary Material for details).  We can prove the asymptotic equivalence between the numerator(denominator) of $\hat{V}_n$ and $V_n$. The result is stated as follows.
	\begin{Th}
		\label{Th4.1}
		Suppose that  Conditions~\ref{CA.1}-\ref{r1} in  Section \ref{sec6} hold.
		Under the null hypothesis in $(\ref{h0})$, if $p^3\log n_1/n_1\to 0$ and $p^3\log n_2/n_2\to 0,$
		\begin{align}
			\hat{V}_n \stackrel{d}{\rightarrow} \mathbf{N}(0,1),
		\end{align}
		where $``\stackrel{d}{\rightarrow} "$ stands for convergence in distribution.
	\end{Th}
	Therefore,  we can compute the critical values easily.
	\begin{remark}
		In Theorem \ref{Th4.1},  we do not restrict the specific relationship between $n_1$ and $n_2$. However, the relationship between $n_1$ and $n_2$  influences the power of our test.  Theorem \ref{Th4.3} below states the results.
	\end{remark}
	\begin{remark}	
		In Theorem \ref{Th4.1}, we require $p^3\log n_1/n_1\to 0$ and $p^3\log n_2/n_2\to 0$ because when analyzing the residuals we  use the asymptotically linear representation of the parameter estimator, and also need the consistency of some sample covariance matrices to  their counterparts at the population level in the sense of $L_2$ norm. \cite{tan2022integrated} showed that to get the asymptotically linear representation of the parameter estimator, this rate cannot be faster in general. However, for linear regression models,  when the rate of divergence can be improved to $p^2=o\left(\max\left\{n_1, n_2\right\}\right)$,  the asymptotically linear representation of the parameter estimator still holds (Theorem 2 in \cite{tan2019adaptive}), and the sample covariance matrices are still consistent.
	\end{remark}
	
	\subsection{Power study}\label{sec3.2}
	Consider the power performance under the alternative hypothesis in (\ref{alter}).  The fixed non-zero constant $\delta_n$ corresponds to
	the global alternative hypothesis with $Y=m(\boldsymbol{X})+\varepsilon $, where $m(\boldsymbol{X})\neq g(\btheta,\boldsymbol{X})$ for $\forall \btheta\in\bTheta$, and   $\delta_n \to 0$ as $n \to \infty$ to  local alternatives.
	Define $\bSigma_{\varepsilon}=E\left\{\varepsilon
	^2\dot{g}(\boldsymbol{\theta}^*,\boldsymbol{X})\dot{g}(\boldsymbol{\theta}^*,\boldsymbol{X})^{\top}\right\},$ $\bSigma_l=E\left\{l^2(\boldsymbol{X})\dot{g}(\boldsymbol{\theta}^*,\boldsymbol{X})\dot{g}(\boldsymbol{\theta}^*,\boldsymbol{X})^{\top}\right\}$ and
	$\bSigma_{cov}^*=E\left\{\dot{g}(\boldsymbol{\theta}^*,\boldsymbol{X}_1)\dot{g}(\boldsymbol{\theta}^*,\boldsymbol{X}_2)^{\top}w_{12}\right\}.$
	We state the following results.
	\begin{Th}
		\label{Th4.3}
		Suppose that Conditions~\ref{CA.1} -- \ref{r3} hold, $p^3\log n_1/n_1\to 0$ and $p^3\log n_2/n_2\to 0$, under the alternative hypothesis in (\ref{alter}),
		
		(a) When $\delta_n=1$ for the global alternatives,  	recalling $l(\boldsymbol{X})=m(\boldsymbol{X})-g(\boldsymbol{\theta}^*,\boldsymbol{X})$, $\bSigma=E\left\{\dot{g}(\btheta^*, \boldsymbol{X}) \dot{g}(\btheta^*, \boldsymbol{X})^{\top}\right\}-E\left[\left\{m(\boldsymbol{X})-g(\btheta^*,\boldsymbol{X})\right\} \ddot{g}(\btheta^*, \boldsymbol{X})\right]$ and $\bSigma_*=E\left\{\dot{g}(\btheta^*, \boldsymbol{X}) \dot{g}(\btheta^*, \boldsymbol{X})^{\top}\right\}$ in Section~\ref{sec2}, we can obtain that
		\begin{align}
			\frac{\hat{V}_n}{\sqrt{n_1}}\stackrel{p}{\to}&\frac{E\left\{l(\boldsymbol{X}_1)l(\boldsymbol{X}_2)w_{12}\right\}}{\sqrt{V_{(0)}}},
		\end{align}
		where $``\stackrel{p}{\to}"$ stands for convergence in probability,  and
		\begin{align*}
			V_{(0)}=&E\left[\left\{\varepsilon_1^2+l^2(\boldsymbol{X}_1)\right\}E_{\boldsymbol{X}_1}^2\left\{l(\boldsymbol{X}_2)w_{12}\right\}\right]-E^2\left\{l(\boldsymbol{X}_1)l(\boldsymbol{X}_2)w_{12}\right\}
			\\&+E\left\{l(\boldsymbol{X}_2)
			\dot{g}(\boldsymbol{\theta}^*, \boldsymbol{X}_1)w_{12}\right\}^{\top}\bSigma_*^{-1}
			\left(\bSigma_{\varepsilon}+\bSigma_{l}\right)
			\bSigma_*^{-1}
			E\left\{l(\boldsymbol{X}_2)
			\dot{g}(\boldsymbol{\theta}^*, \boldsymbol{X}_1)w_{12}\right\}
			\\&-2E\left[\left\{\varepsilon_1^2+l^2(\boldsymbol{X}_1)\right\}l(\boldsymbol{X}_2)
			\dot{g}(\boldsymbol{\theta}^*, \boldsymbol{X}_1)w_{12}\right]^{\top}\bSigma_*^{-1}
			E\left\{l(\boldsymbol{X}_2)\dot{g}(\boldsymbol{\theta}^*, \boldsymbol{X}_1)w_{12}\right\}
			.
		\end{align*}
		In cases (b)-(d), recalling $l(\boldsymbol{X})=m(\boldsymbol{X})-g(\boldsymbol{\theta}_0,\boldsymbol{X})$ and $\bSigma=E\left\{\dot{g}(\btheta_0, \boldsymbol{X}) \dot{g}(\btheta_0, \boldsymbol{X})^{\top}\right\}$, we have:
		
		(b) When $\delta_n=1/\sqrt{n}$,
		\begin{align}		\hat{V}_n-\frac{\frac{\sqrt{n_1}}{\sqrt{n_2n}}\sum_{j=n_1+1}^{n}\varepsilon_jH(\boldsymbol{X}_j)+\frac{\sqrt{n_1n_2}}{n}\mu}{\sqrt{E_{\mathcal{N}_2}\left[\left\{\frac{1}{\sqrt{n_2}}\sum_{j=n_1+1}^{n}\varepsilon_1\varepsilon_jw{'}_{1j}+\frac{\sqrt{n_2}}{\sqrt{n}}\varepsilon_1H(\boldsymbol{X}_1)\right\}^2\right]}}\stackrel{d}{\to}\mathbf{N}(0,1).
		\end{align}
		where
		\begin{align*}
			w{'}_{1j}=&w_{1j}-\dot{g}(\boldsymbol{\theta}^*,\boldsymbol{X}_j)^{\top}\bSigma^{-1}E_{\boldsymbol{X}_1}\left\{\dot{g}(\boldsymbol{\theta}^*,\boldsymbol{X}_j)w_{1j}\right\}-\dot{g}(\boldsymbol{\theta}^*, \boldsymbol{X}_{1})^{\top}\bSigma^{-1}E_{\boldsymbol{X}_j}\left\{\dot{g}(\boldsymbol{\theta}^*,\boldsymbol{X}_1)w_{1j}\right\}\nonumber\\&+\dot{g}(\boldsymbol{\theta}^*, \boldsymbol{X}_{1})^{\top}\bSigma^{-1}\bSigma_{cov}^*\bSigma^{-1}\dot{g}(\boldsymbol{\theta}^*,\boldsymbol{X}_j),
		\end{align*}
		\begin{align*} \mu=&E\left\{l(\boldsymbol{X}_1)l(\boldsymbol{X}_2)w_{12}\right\} -2E\left\{\dot{g}(\boldsymbol{\theta}_0,\boldsymbol{X}_1)l(\boldsymbol{X}_2)w_{12}\right\}^{\top}\bSigma^{-1}E\left\{\dot{g}(\boldsymbol{\theta}_0,\boldsymbol{X}_1)l(\boldsymbol{X}_1)\right\} \\&+E\left\{\dot{g}(\boldsymbol{\theta}_0,\boldsymbol{X}_1)l(\boldsymbol{X}_1)\right\}^{\top}\bSigma^{-1}\bSigma_{cov}^*\bSigma^{-1}E\left\{\dot{g}(\boldsymbol{\theta}_0,\boldsymbol{X}_1)l(\boldsymbol{X}_1)\right\},
		\end{align*}
		\begin{align*}
			H(\boldsymbol{X}_1)=& E_{\boldsymbol{X}_1}\left\{l(\boldsymbol{X}_2)w_{12}\right\}- E_{\boldsymbol{X}_1}\left\{\dot{g}(\boldsymbol{\theta}_0,\boldsymbol{X}_2)w_{12}\right\}^{\top}\bSigma^{-1}E\left\{\dot{g}(\boldsymbol{\theta}_0,\boldsymbol{X}_1)l(\boldsymbol{X}_1)\right\}\nonumber
			\\&-\dot{g}(\boldsymbol{\theta}_0, \boldsymbol{X}_{1})^{\top} \bSigma^{-1}E\left\{\dot{g}(\boldsymbol{\theta}_0,\boldsymbol{X}_1) l(\boldsymbol{X}_2)w_{12}\right\}\nonumber
			\\&+\dot{g}(\boldsymbol{\theta}_0, \boldsymbol{X}_{1})^{\top}\bSigma^{-1}\bSigma_{cov}^*\bSigma^{-1}E\left\{\dot{g}(\boldsymbol{\theta}_0,\boldsymbol{X}_1)l(\boldsymbol{X}_1)\right\},
		\end{align*}
		and $\sum_{j=n_1+1}^{n}\varepsilon_jH(\boldsymbol{X}_j)/\sqrt{n_2}$ converges to $\mathbf{N}\left(0,E\left[\left\{ \varepsilon_1H(\boldsymbol{X}_1)\right\}^2\right]\right)$ in distribution.
		
		Especially, if $n_1/n=o(1)$,
		then $	\hat{V}_n\stackrel{d}{\to}
		\mathbf{N}(0,1). $
		
		(c) When $\delta_n=n^{-\alpha}$,  $\frac{1}{4}<\alpha<\frac{1}{2}$, $n^{1-4\alpha}p^3\log n\to 0$, $n^{-\alpha}\sqrt{n_1}\to \infty$, then $ \hat{V}_n\stackrel{p}\to\infty. $
		
		(d) When $\delta_n=n^{-\alpha}$, $\alpha>\frac{1}{2}$, then $	\hat{V}_n\stackrel{d}{\to}
		\mathbf{N}(0,1).  $
		
	\end{Th}
	\begin{remark}
		Theorem \ref{Th4.3} shows that the test can detect the local alternatives distinct from the null at the fastest possible rate of order $1/\sqrt {n_1}$ in general. Due to possible different sizes $n_1$ and $n_2$ of the two subsamples, the above analysis presents  more detailed results in different cases than those with existing tests in the literature. From the above results, we can see that  only in case (b) with $\delta_n=1/\sqrt n$, the optimal splitting is $n_1=n_2=n/2$, while in cases (a) and (c), large $n_1$ can enhance power.  But practically, if $n_2$ is too small, the conditional variance cannot be estimated well causing that the test may not perform well. These claims were confirmed when we conducted numerical studies using  $n_2=0.5n$, $n_2=0.25n$, and $n_2=0.1n$.  Thus, in the numerical studies, we report the results with $n_2=0.25n$.  Another issue is the power performance when $\delta_n=n^{-\alpha}$ with $0< \alpha<1/4$. We do not discuss this case mainly because of the difficulty of studying the negligibility of the remaining terms in the asymptotically linear representations of $\boldsymbol{\hat{\theta}} _1-\boldsymbol{\theta}_0$ and $\boldsymbol{\hat \theta}_2-\boldsymbol{\theta}_0$ under the local alternatives.
	\end{remark}
	\section{Numerical Studies}\label{sec4}
	\subsection{Simulations}
	\par In this section, some numerical studies are conducted to examine the performance of our test proposed in Section 3. As \cite{li2019model} used, the weight function $1/\sqrt {\|\boldsymbol{X}_1-\boldsymbol{X}_2\|^2+1}$ has the merits that combine local smoothing and global smoothing tests. But it is theoretically flawed for large $p$ as it converges to a constant when $p$ goes to infinity. To remedy this defect, not only should our chosen weight function include this weight function, but also it contains another weight function  $\sum_{k=1}^{p}K\left\{(X_{ik}-X_{jk})/h\right\}/h $, where $K(\cdot)$ denotes the kernel density function and $h$ is the bandwidth. The summation form ensures that it works in diverging dimension cases. As a result, the weight function, defined as $0.5\times \left[1/\sqrt{\|\boldsymbol{X}_1-\boldsymbol{X}_2\|^2+1}+ \sum_{k=1}^{p}K\left\{(X_{ik}-X_{jk})/h\right\}/h\right]$, is a hybrid of two equally weighted functions.
	
\begin{figure}[H]
		\centering
		\includegraphics[scale=0.45]{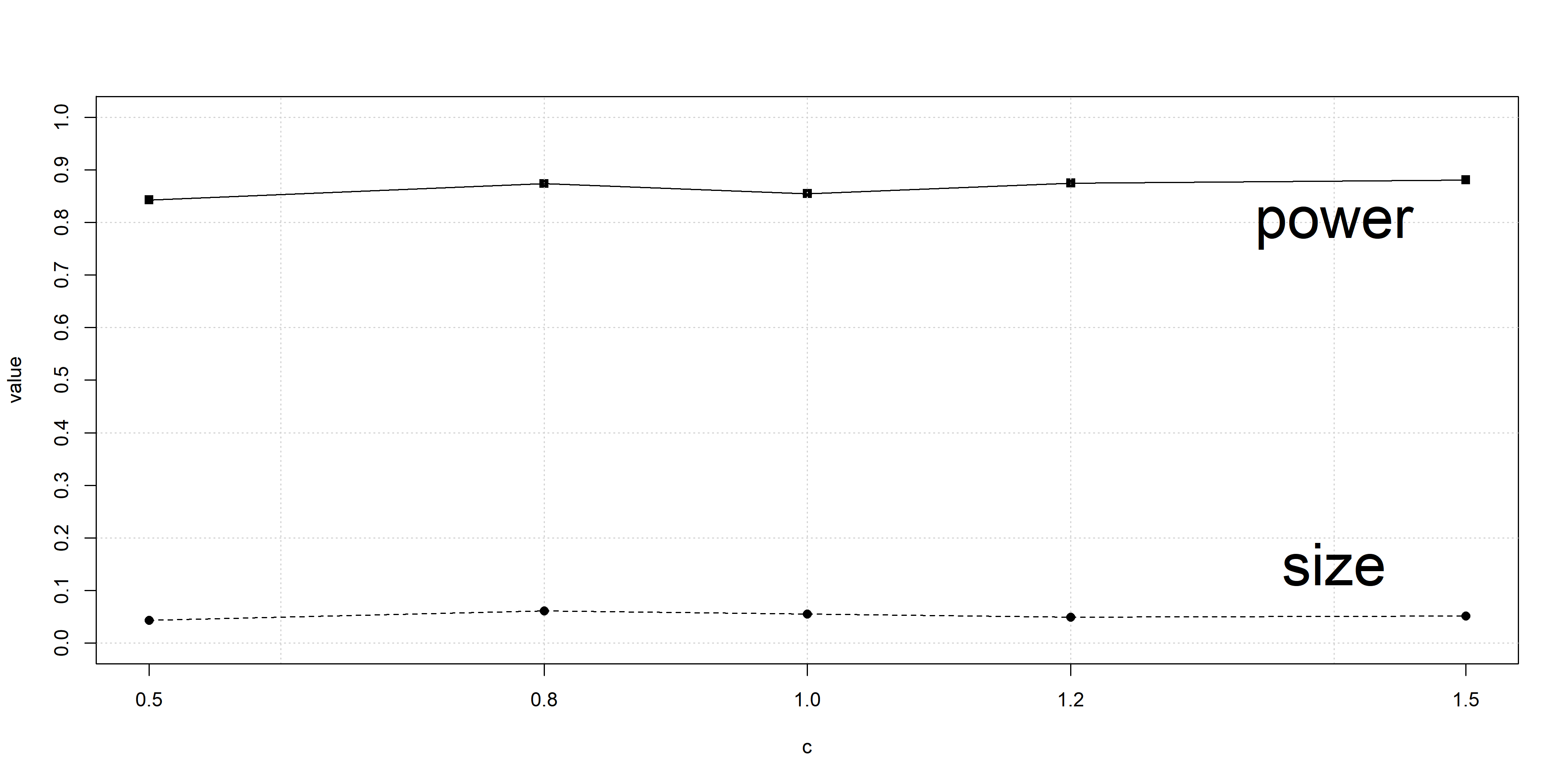}
		\caption{Empirical sizes and powers of $Cost_n$ under different bandwidths for $H_{11}$ in Study 1, when $n=400$, $q=17$ and $\bSigma=\bSigma_1 $ .}
		\label{P1}
	\end{figure}

	To make the simulation results convincing, we compare with the test $AICM_n$ proposed by \cite{tan2022integrated} that can also handle diverging dimension cases and shows its advantages.  It is worth noticing that the method in \cite{jankova2020goodness} could also be applied to non-sparse models with random designs and dimensions under our constraints,   \cite{tan2022integrated}  made a comparison for Logistic model with three model settings, and $AICM_n$ outperformed the test in \cite{jankova2020goodness}  in two  of the three model settings. We have also conducted the simulations under the same model settings and found that $COST$ worked similarly to $AICM_n$. Therefore, we put the results in Section 4.2,  and the main text here only reports the numerical comparison with  $AICM_n$.  In addition, we also evaluate the performance of another related test, $DrCost_n$, which can be seen as a modified version of  $Cost_n$. When the underlying model has a dimension reduction structure, $DrCost_n$ contains the dimension reduction step (see, \cite{tan2022integrated}) .
	We design three studies: models with dimension reduction structures;  with dimension reduction structures and diverging dimensions; and without dimension reduction structures. The first two studies favor \cite{tan2022integrated}'s method, and the third study deals with general models. The predictor vectors $\boldsymbol{X}_i$'s are independently generated from multivariate normal distribution $\mathbf{N}(\boldsymbol{0},\bSigma) $. Here $\bSigma=\bSigma_1=I_p$ or $\bSigma= \bSigma_2=(0.5^{|i-j|})_{p\times p}$. The error $\varepsilon$'s are independently drawn from the standard
	normal distribution $\mathbf{N}(0,1)$. The simulation results are all based on $1,000$ replications. The results of $AICM_n$ are computed by the code provided by the authors of the paper \cite{tan2022integrated}. To choose the bandwidth $h$, we consider it to be $c\cdot n^{-0.2} $, where $c$ is a constant taking the following five values: $0.5,0.8,1,1.2,1.5$. To check how robust the test is against the value $c$, we use the model $H_{11}$ as an example in Figure \ref{P1}. It shows that our test performs robustly, and the size and power level can be similar. Therefore,  we use the bandwidth with $c=1$.
	
	For comparisons, we consider four numerical studies. Studies~1 and 2 have multi-index model structures that favor $AICM_n$, and Study~3 does not have such a structure. As under some strong conditions on the regression function our method can handle the cases where  the dimension $q$ of the predictor vector can be much higher than that $p$ of parameter vector, we consider Study~4 that has $q=p^2$ and $q=n$. As the limiting null distribution of $AICM_n$ is generally intractable, the wild bootstrap approximation is used to determine the critical values. 
	For model $H_{11}$ in Study~1 and model $H_{21}$ in Study~2, the numerical results of $AICM_n$ are excerpted from \cite{tan2022integrated} to make the paper self-contained.
	
	{\it Study 1.}  Generate data from the following double-index and triple-index models:
	\begin{align*}
		&H_{11}: Y=\bbeta_{1}^{\top} \boldsymbol{X}+a\left(\bbeta_{2}^{\top} \boldsymbol{X}\right)^{2}+\varepsilon, \\
		&H_{12}: Y=X_1+\cos(2X_2)+a\exp(3X_2) +\varepsilon.
	\end{align*}
	We set $\bbeta_1=(\underbrace{1,\cdots,1}_{q_1},0,\cdots,0)^{\top}/\sqrt{q_1}$ and
	$\bbeta_2=(0,\cdots,0,\underbrace{1,\cdots,1}_{q_1})^{\top}/\sqrt{q_1}$ with $q_1= [q/2]_{-} $, where $ [ \quad]_{-} $ indicates the largest integer smaller than or equal to $q/2$. Here $X_i $ denotes the $i$-th component of $ \boldsymbol{X} $.  The first  hypothetical model is single-index, and the second is double-index, containing the second index and third index in alternative models respectively.
	The empirical sizes and powers are reported in Tables~\ref{T11}-\ref{T12} in Section~\ref{section4.2}.
	
	Table \ref{T11} shows that $AICM_n$  works better for model $H_{11}$, and $DrCost_n$ with the dimension reduction step  outperforms $Cost_n$, but slightly. When the sample size gets larger, our tests gradually work closer to $AICM_n$.  However, for model $H_{12}$, the situation changes. The results of Table~\ref{T12} suggest that  $Cost_n$ outperforms  $DrCost_n$ and $AICM_n$, although the model favors them. We checked the details and found that the structural dimension of the central subspace is underestimated to $1$ in this model and the residuals cannot be estimated well under the alternative hypothesis. This might be one of the reasons that $AICM_n$ and $DrCost_n$ do not work well.
	
	{\it Study 2.}  Generate data from the following multi-index models with $q=0.1 n$ and $q=\sqrt n$ respectively:
	\begin{align*}
		&H_{21}: Y=\bbeta_{0}^{\top} \boldsymbol{X}+a \exp \left(\bbeta_{0}^{\top} \boldsymbol{X}\right)+\varepsilon, \\
		&H_{22}: Y=\bbeta_{1}^{\top} \boldsymbol{X}+\exp \left(\bbeta_{2}^{\top} \boldsymbol{X}\right)+a \exp \left(-\bbeta_{0}^{\top} \boldsymbol{X}\right)+\varepsilon,
	\end{align*}
	where $\bbeta_0=(1,1,\cdots,1)^{\top}/\sqrt{q}$ and other notations are the same as stated above.  In this study, the dimension $q$ is large; thus, the regularity conditions fail to hold in theory. The empirical sizes and powers of Study 2 are presented in Tables \ref{T21}-\ref{T22} in Section~\ref{section4.2}.
	
	 The simulation results suggest that all competitors cannot maintain the significance level for model $H_{21} $ when $q$ diverges at the rate of $0.1n$. This means that the condition of dimensionality is violated too much. For model $H_{22}$ with  $q=\sqrt{n}$ that also violates the condition,  we can see that when the sample size is large, our test performs well in the significance level maintenance with relatively high power. At the same time, $AICM_n$  is liberal in general. The test $DrCost_n$ with the dimension reduction step works slightly worse than $Cost_n$. This suggests that the test may still be usable when the sample size is large.
	
{\it Study 3.} Generate data from the following models without dimension reduction structures:
	\begin{align*}
		&H_{31}: Y=X_1+\cos(2X_2)+a\sum_{i=1}^{q}\exp(3X_i) +\varepsilon, \\
		&H_{32}: Y=\sum_{i=1}^{q}\sin(\beta_{0i}X_i)+a\sum_{i=1}^{q}\exp(3X_i) +\varepsilon,\\
		&H_{33}: Y=\sum_{i=1}^{q-1}X_{i}X_{i+1}+a\cos\left(\bbeta_{0}^{\top} \boldsymbol{X}\right)+\varepsilon,\\
		&H_{34}: Y=\sum_{i=1}^{q-2}X_{i}X_{i+1}\sin(\pi X_{i+2})+a\left(\bbeta_{0}^{\top} \boldsymbol{X}\right)^3+\varepsilon,
	\end{align*}
	where $\beta_{0i} $ denotes the $i$-th component of $ \bbeta_{0} $ and the rest of the notations are the same as stated above.  The four models have no dimension reduction structures, as the representatives of models: under the null,
	with low-dimensional and high-dimensional regressions; under the alternatives, with high-dimensional
	departure; and under the null, with high-dimensional regressions with interactions among the covariates.
	
	Tables \ref{T31}-\ref{T34} in Section~\ref{section4.2}
	report the empirical sizes and powers. $AICM_n$ fails to work entirely in high dimension scenarios, especially for Model $H_{34}$ allowing higher order interactions between the covariates. In almost all cases, $AICM_n$ cannot maintain the significance level and even has no empirical powers. This phenomenon suggests that $AICM_n$ relies critically on dimension reduction structures.  The test $Cost_n$ works much better comparably as expected.
	
		{\it Study 4.} Generate data from the following models with $q=p^2$ and $q=n $ respectively:
		\begin{align*}
			&H_{41}: Y=\sum_{i=1}^{p}\sin\left(\beta_{1i}\prod_{j=(i-1)*r+1}^{\min(i*r,q)}X_j\right)+a\left(\bbeta_1^\top \boldsymbol{X}\right)^2 +\varepsilon, \\
			&H_{42}: Y=\sum_{i=1}^{p}\sin\left(\beta_{1i}\sum_{j=(i-1)*r+1}^{(i-1)*r+r_1}X_j+\sum_{j=(i-1)*r+r_1+1}^{\min(i*r,q)}X_j\right)+a\left(\bbeta_1^\top \boldsymbol{X}\right)^2  +\varepsilon,
		\end{align*}
		where $ \boldsymbol{X}=(X_1,X_2,\cdots,X_p)^\top$, $\bbeta_1=(\underbrace{1,\cdots,1}_{p_1},0,\cdots,0)^{\top}/\sqrt{p_1}$ with $p_1= [p/2]_{-} $ and $\beta_{1i} $ denotes the $i$-th component of $ \bbeta_1 $. In addition, define $r=[q/p]^{+}$ and $r_1=[r/2]_{-}$, where $[\quad ]^{+}$ takes the smallest integer greater than or equal to $q/p $. The other notations remain the same as mentioned earlier. The two models have much higher  dimensions of the predictor vector than that of the parameter vector.  
		Tables \ref{T41}-\ref{T42} in Section~\ref{section4.2}
		report the empirical sizes and powers. The simulation results suggest that $Cost_n$  may not be significantly  affected by the dimension $q$ of the predictor vector when the regression function has a particular structure about the predictors and  still works well in both significance level maintenance and power performance, whereas $AICM_n$  entirely fails to work.

	The performance of our test is more robust against model settings than $AICM_n$. Thus, our test is more suitable when the sample size is relatively large. But in moderate sample size scenarios, the test loses some power. See Table~7 for instance.  This is because  the sample-splitting technique causes  the sample size reduction such that the test has a slower rate to the weak limit than the classic tests. Another reason could be because of using the limiting null distribution to determine the critical values whereas $AICM_n$ uses the bootstrap approximation to favor the small and moderate sample size scenarios.
	
	\subsection{Simulation results}
	\label{section4.2}
	\begin{table}[H]
			\renewcommand\arraystretch{0.55}
		\caption{Empirical sizes and powers for $H_{11}$ in Study 1.}
	    \center{
			\begin{tabular}{ccccccccc}
				\hline
				~ & \multirow{3}*{a} & n=100 & n=100 & n=100& n=100 & n=200 & n=400 & n=600 \\
				~ & ~ & q=2 & q=4 & q=6 & q=8 & q=12 & q=17 & q=20 \\
				~ & ~ & ~ & ~ & ~ & p=q & ~ & ~ & ~ \\ \hline
				$Cost_n,\bSigma_1$ & 0.00  & 0.044 & 0.044 & 0.048 & 0.055 & 0.061 & 0.045 & 0.053  \\
				~ & 0.25  & 0.326 & 0.313 & 0.367 & 0.357 & 0.674 & 0.951 & 0.995  \\ \hline
				$DrCost_n,\bSigma_1$ & 0.00  & 0.050 & 0.046 & 0.042 & 0.054 & 0.050 & 0.041 & 0.048  \\
				~ & 0.25  & 0.343 & 0.348 & 0.339 & 0.304 & 0.620 & 0.917 & 0.979  \\ \hline
				$AICM_n,\bSigma_1$&0.00& 0.055& 0.051& 0.076& 0.051& 0.050& 0.050& 0.065\\
				(from \cite{tan2022integrated})&0.25& 0.556& 0.564& 0.553& 0.562& 0.853& 0.992& 1.000\\
				\hline
				\hline
				$Cost_n,\bSigma_2$ & 0.00  & 0.043 & 0.053 & 0.052 & 0.061 & 0.048 & 0.054 & 0.055  \\
				~ & 0.25  & 0.337 & 0.624 & 0.767 & 0.817 & 0.996 & 1.000 & 1.000  \\ \hline
				$DrCost_n,\bSigma_2$ & 0.00  & 0.051 & 0.049 & 0.046 & 0.056 & 0.058 & 0.048 & 0.065  \\
				~ & 0.25  & 0.281 & 0.540 & 0.703 & 0.768 & 0.987 & 1.000 & 1.000  \\ \hline
				$AICM_n,\bSigma_2$&0.00& 0.052& 0.043& 0.059& 0.070& 0.057& 0.049& 0.050\\
				(from \cite{tan2022integrated})&0.25& 0.481& 0.820& 0.916& 0.956& 1.000& 1.000& 1.000\\
				\hline
			\end{tabular}
			\label{T11}	}
	\end{table}
	
	\begin{table}[H]
		\renewcommand\arraystretch{0.55}
		\caption{Empirical sizes and powers for $H_{12}$ in Study 1.}
		\center{
			\begin{tabular}{ccccccccc}
				\hline
				~ & \multirow{3}*{a} & n=100 & n=100 & n=100& n=100 & n=200 & n=400 & n=600 \\
				~ & ~ & q=2 & q=4 & q=6 & q=8 & q=12 & q=17 & q=20 \\
				~ & ~ & ~ & ~ & ~ &  p=2 & ~ & ~ & ~ \\ \hline
				$Cost_n,\bSigma_1$ & 0.00 &0.054 & 0.048 & 0.038 & 0.055 & 0.046 & 0.045 & 0.050 \\
				~ & 0.10 & 0.530 & 0.487 & 0.473 & 0.533 & 0.726 & 0.878 & 0.923 \\ \hline
				$DrCost_n,\bSigma_1$ & 0.00  &  0.051 & 0.051 & 0.060 & 0.042 & 0.054 & 0.054 & 0.046\\
				~ & 0.10  &  0.403 & 0.425 & 0.415 & 0.390 & 0.619 & 0.757 & 0.850 \\ \hline
				$AICM_n,\bSigma_1$ & 0.00 &  0.054  & 0.059  & 0.044  & 0.051  & 0.069  & 0.066  & 0.068\\
				~ & 0.10 & 0.098  & 0.093  & 0.084  & 0.103  & 0.112  & 0.172  & 0.240 \\ \hline
				\hline
				$Cost_n,\bSigma_2$ & 0.00 &  0.051 & 0.036 & 0.052 & 0.043 & 0.038 & 0.039 & 0.039\\
				~ & 0.10 &0.502 & 0.509 & 0.557 & 0.493 & 0.743 & 0.830 & 0.885 \\ \hline
				$DrCost_n,\bSigma_2$ & 0.00  & 0.050 & 0.045 & 0.044 & 0.040 & 0.046 & 0.053 & 0.052\\
				~ & 0.10  &  0.248 & 0.266 & 0.288 & 0.294 & 0.457 & 0.601 & 0.688 \\ \hline
				$AICM_n,\bSigma_2$&0.00&  0.050  & 0.053  & 0.048  & 0.038  & 0.059  & 0.064  & 0.053  \\
				~&0.10& 0.043  & 0.063  & 0.050  & 0.051  & 0.042  & 0.031  & 0.018\\
				\hline
			\end{tabular}
			\label{T12}	}
	\end{table}

\begin{table}[H]
	\renewcommand\arraystretch{0.55}
	\caption{Empirical sizes and powers for $H_{21}$ in Study 2 with $q=0.1n$.}
	\center{
		\begin{tabular}{cccccc}
			\hline
			~ & \multirow{3}*{a} & n=50 & n=100  & n=500 & n=1000\\
			~ & ~ & q=5 & q=10 & q=50 & q=100\\
			~ & ~ & ~ &p=q & ~& ~\\ \hline
			$Cost_n,\bSigma_1$ & 0.00 & 0.039 & 0.058 & 0.063 & 0.064  \\
			~ & 0.10 & 0.116 & 0.222 & 0.798 & 0.984  \\ \hline
			$DrCost_n,\bSigma_1$ & 0.00  & 0.067 & 0.053 & 0.065 & 0.075  \\
			~ & 0.10  & 0.107 & 0.151 & 0.566 & 0.871 \\ \hline
			$AICM_n,\bSigma_1$&0.00& 0.062& 0.057&  0.071& 0.081\\
			(from\cite{tan2022integrated})&0.10& 0.163& 0.250& 0.858& 0.994\\
			\hline
			\hline
			$Cost_n,\bSigma_2$ & 0.00 & 0.051 & 0.057 & 0.058 & 0.072  \\
			~ & 0.10 & 0.184 & 0.461 & 0.993 & 1.000\\
			\hline
			$DrCost_n,\bSigma_2$ & 0.00  & 0.050 & 0.059 & 0.070 & 0.071  \\
			~ & 0.10  & 0.154 & 0.331 & 0.970 & 0.992  \\ \hline
			$AICM_n,\bSigma_2$&0.00& 0.064& 0.068&  0.079& 0.107\\
			(from \cite{tan2022integrated})&0.10& 0.235& 0.582 & 0.935& 0.959\\
			\hline
		\end{tabular}
		\label{T21}	}
\end{table}

\begin{table}[H]
	\renewcommand\arraystretch{0.55}
	\caption{Empirical sizes and powers for $H_{22}$ in Study 2 with $q=\sqrt n$.}
	\center{
		\begin{tabular}{ccccc}
			\hline
			~ & \multirow{3}*{a} & n=100 &  n=400 & n=900 \\
			~ & ~ & q=10  & q=20 & q=30 \\
			~ & ~ & ~& p=2q  & ~\\ \hline
			$Cost_n,\bSigma_1$ & 0.00 & 0.104 & 0.068 & 0.061  \\
			~ & 0.50 & 0.568 & 0.991 & 1.000  \\ \hline
			$DrCost_n,\bSigma_1$ & 0.00  & 0.094 & 0.060 & 0.059  \\
			~ & 0.50  & 0.512 & 0.962 & 1.000 \\
			\hline
			$AICM_n,\bSigma_1$ & 0.00  & 0.079   & 0.074  & 0.074  \\
			~ & 0.50  & 0.970   & 0.999  & 1.000  \\
			\hline
			\hline
			$Cost_n,\bSigma_2$ & 0.00 & 0.056 & 0.059 & 0.037   \\
			~ & 0.50 & 0.393 & 0.824 & 0.953 \\ \hline
			$DrCost_n,\bSigma_2$ & 0.00  & 0.074 & 0.061 & 0.057  \\
			~ & 0.50  & 0.358 & 0.708 & 0.889  \\ \hline
			$AICM_n,\bSigma_2$ & 0.00  & 0.086   & 0.091  & 0.069  \\
			~ & 0.50  & 0.760  & 0.839  & 0.917  \\
			\hline
		\end{tabular}
		\label{T22}	}
\end{table}

	\begin{table}[H]
	\renewcommand\arraystretch{0.55}
	\caption{Empirical sizes and powers for $H_{31}$ in Study 3.}
	\center{
		\begin{tabular}{ccccccccc}
			\hline
			~ & \multirow{3}*{a} & n=100 & n=100 & n=100& n=100 & n=200 & n=400 & n=600 \\
			~ & ~ & q=2 & q=4 & q=6 & q=8 & q=12 & q=17 & q=20 \\
			~ & ~ & ~ & ~ & ~ &  p=2 & ~ & ~ & ~ \\ \hline
			$Cost_n,\bSigma_1$ & 0.00 &  0.042 & 0.035 & 0.043 & 0.046 & 0.057 & 0.048 & 0.047 \\
			~ & 0.10 & 0.658 & 0.726 & 0.825 & 0.817 & 0.933 & 0.975 & 0.970\\ \hline
			$AICM_n,\bSigma_1$ & 0.00 & 0.049  & 0.057  & 0.050  & 0.048  & 0.069  & 0.065  & 0.066 \\
			~ & 0.10 & 0.041  & 0.033  & 0.063  & 0.082  & 0.144  & 0.211  & 0.258 \\ \hline
			\hline
			$Cost_n,\bSigma_2$ & 0.00 & 0.034 & 0.045 & 0.041 & 0.045 & 0.039 & 0.048 & 0.051  \\
			~ & 0.10 & 0.610 & 0.738 & 0.850 & 0.838 & 0.935 & 0.950 & 0.978 \\ \hline
			$AICM_n,\bSigma_2$ & 0.00 & 0.050  & 0.055  & 0.048  & 0.039  & 0.061  & 0.063  & 0.056  \\
			~ & 0.10 & 0.028  & 0.010  & 0.017  & 0.022  & 0.041  & 0.070  & 0.104 \\ \hline
		\end{tabular}
		\label{T31}	}
\end{table}

\begin{table}[H]
	\renewcommand\arraystretch{0.55}
	\caption{Empirical sizes and powers for $H_{32}$ in Study 3.}
	\center{
		\begin{tabular}{ccccccccc}
			\hline
			~ & \multirow{3}*{a} & n=100 & n=100 & n=100& n=100 & n=200 & n=400 & n=600 \\
			~ & ~ & q=2 & q=4 & q=6 & q=8 & q=12 & q=17 & q=20 \\
			~ & ~ & ~ & ~ & ~ &  p=q & ~ & ~ & ~ \\ \hline
			$Cost_n,\bSigma_1$ & 0.00 &  0.035 & 0.029 & 0.046 & 0.046 & 0.055 & 0.064 & 0.041   \\
			~ & 0.10 & 0.588 & 0.757 & 0.782 & 0.840 & 0.988 & 1.000 & 1.000\\ \hline
			$AICM_n,\bSigma_1$ & 0.00 & 0.061 & 0.056 & 0.056 & 0.079 & 0.074 & 0.048 & 0.000 \\
			~ & 0.10 & 0.353 & 0.162 & 0.03 & 0.002 & 0.000 & 0.000 & 0.000 \\ \hline
			\hline
			$Cost_n,\bSigma_2$ & 0.00 & 0.042 & 0.051 & 0.053 & 0.051 & 0.051 & 0.070 & 0.049   \\
			~ & 0.10 & 0.487 & 0.637 & 0.760 & 0.843 & 0.967 & 0.998 & 1.000 \\ \hline
			$AICM_n,\bSigma_2$ & 0.00 & 0.060 & 0.063 & 0.047 & 0.057 & 0.064 & 0.053 & 0.000  \\
			~ & 0.10 & 0.391 & 0.177 & 0.022 & 0.000 & 0.000 & 0.000 & 0.000 \\ \hline
		\end{tabular}
		\label{T32}	}
\end{table}

\begin{table}[H]
	\renewcommand\arraystretch{0.55}
	\caption{Empirical sizes and powers for $H_{33}$ in Study 3.}
	\center{
		\begin{tabular}{ccccccccc}
			\hline
			~ & \multirow{3}*{a} & n=100 & n=100 & n=100& n=100 & n=200 & n=400 & n=600 \\
			~ & ~ & q=2 & q=4 & q=6 & q=8 & q=12 & q=17 & q=20 \\
			~ & ~ & ~ & ~ & ~ &  p=q-1 & ~ & ~ & ~ \\ \hline
			$Cost_n,\bSigma_1$ & 0.00 & 0.047 & 0.054 & 0.062 & 0.042 & 0.058 & 0.039 & 0.052  \\
			~ & 0.50 & 0.688 & 0.647 & 0.644 & 0.615 & 0.915 & 0.996 & 0.998 \\ \hline
			$AICM_n,\bSigma_1$ & 0.00 & 0.052  & 0.038  & 0.020  & 0.002  & 0.000  & 0.000  & 0.000  \\
			~ & 0.50 & 0.905  & 0.792  & 0.550  & 0.144  & 0.000  & 0.000  & 0.000  \\ \hline
			\hline
			$Cost_n,\bSigma_2$ & 0.00 & 0.035 & 0.052 & 0.051 & 0.048 & 0.044 & 0.053 & 0.051  \\
			~ & 0.50 & 0.675 & 0.558 & 0.465 & 0.399 & 0.572 & 0.774 & 0.867   \\ \hline
			$AICM_n,\bSigma_2$ & 0.00 & 0.060  & 0.038  & 0.025  & 0.004  & 0.000  & 0.000  & 0.000  \\
			~ & 0.50 & 0.904  & 0.724  & 0.496  & 0.162  & 0.000  & 0.000  & 0.000 \\ \hline
		\end{tabular}
		\label{T33}	}
\end{table}

\begin{table}[H]
	\renewcommand\arraystretch{0.55}
	\caption{Empirical sizes and powers for $H_{34}$ in Study 3.}
	\center{
		\begin{tabular}{ccccccccc}
			\hline
			~ & \multirow{3}*{a} & n=100 & n=100 & n=100& n=100 & n=200 & n=400 & n=600 \\
			~ & ~ & q=2 & q=4 & q=6 & q=8 & q=12 & q=17 & q=20 \\
			~ & ~ & ~ & ~ & ~ &  p=q-2 & ~ & ~ & ~ \\ \hline
			$Cost_n,\bSigma_1$ & 0.00 & / & 0.052  & 0.047  & 0.046  & 0.053  & 0.057  & 0.055 \\
			~ & 0.50 & / & 0.489  & 0.345  & 0.226  & 0.433  & 0.661  & 0.787 \\ \hline
			$AICM_n,\bSigma_1$ & 0.00 & / & 0.038  & 0.015  & 0.000  & 0.000  & 0.000  & 0.000  \\
			~ & 0.50 & / & 0.642  & 0.176  & 0.022  & 0.000  & 0.000  & 0.000 \\ \hline
			\hline
			$Cost_n,\bSigma_2$ & 0.00 & / & 0.048  & 0.044  & 0.057  & 0.051  & 0.042  & 0.066  \\
			~ & 0.50 & / & 0.909  & 0.835  & 0.750  & 0.932  & 0.995  & 0.999  \\ \hline
			$AICM_n,\bSigma_2$ & 0.00 & / & 0.047  & 0.022  & 0.005  & 0.000  & 0.000  & 0.000 \\
			~ & 0.50 & / & 0.619  & 0.251  & 0.075  & 0.032  & 0.004  & 0.000 \\ \hline
		\end{tabular}
		\label{T34}	}
\end{table}

	\begin{table}[H]
	\renewcommand\arraystretch{0.55}
	\caption{Empirical sizes and powers for $H_{41}$ in Study 4.}
	\center{
		\begin{tabular}{ccccccccc}
			\hline
			~ & \multirow{2}*{a} & n=100 & n=100 & n=100& n=100 & n=200 & n=400 & n=600 \\
			~ & ~ & p=2 & p=4 & p=6 & p=8 & p=12 & p=17 & p=20 \\ \hline
			~ & ~ & ~ & ~ & $q=p^2$ & ~ & ~ & ~ & ~ \\ \hline
			$Cost_n,\bSigma_1$ & 0.00 & 0.053  & 0.063  & 0.047  & 0.048  & 0.046  & 0.055  & 0.049 \\
			~ & 0.25 &0.357  & 0.399  & 0.421  & 0.460  & 0.762  & 0.965  & 0.994 \\ \hline
			$AICM_n,\bSigma_1$ & 0.00 & 0.037  & 0.000  & 0.000  & 0.000  & 0.000  & 0.000  & 0.000  \\
			~ & 0.25 & 0.279  & 0.000  & 0.000  & 0.000  & 0.000  & 0.000  & 0.000  \\ \hline
			\hline
			$Cost_n,\bSigma_2$ & 0.00 & 0.044  & 0.052  & 0.051  & 0.047  & 0.055  & 0.046  & 0.039   \\
			~ & 0.25 & 0.272  & 0.645  & 0.841  & 0.913  & 1.000  & 1.000  & 1.000  \\ \hline
			$AICM_n,\bSigma_2$ & 0.00 & 0.094  &  0.000  & 0.000  & 0.000  & 0.000  & 0.000  & 0.000   \\
			~ & 0.25 & 0.206 &  0.000  & 0.000  & 0.000  & 0.000  & 0.000  & 0.000  \\ \hline
			~ & ~ & ~ & ~ & $q=n$ & ~ & ~ & ~ & ~\\ \hline
			$Cost_n,\bSigma_1$ & 0.00 & 0.037  & 0.041  & 0.049  & 0.043  & 0.051  & 0.059  & 0.041 \\
			~ & 0.25 & 0.486  & 0.424  & 0.432  & 0.435  & 0.765  & 0.961  & 1.000 \\ \hline
			$AICM_n,\bSigma_1$ & 0.00 & 0.000  & 0.000  & 0.000  & 0.000  & 0.000  & 0.000  & 0.000   \\
			~ & 0.25 &0.000  & 0.000  & 0.000  & 0.000  & 0.000  & 0.000  & 0.000  \\ \hline
			\hline
			$Cost_n,\bSigma_2$ & 0.00 & 0.047  & 0.052  & 0.037  & 0.041  & 0.063  & 0.039  & 0.065  \\
			~ & 0.25 & 0.455  & 0.766  & 0.881  & 0.921  & 0.994  & 1.000  & 1.000  \\ \hline
			$AICM_n,\bSigma_2$ & 0.00 & 0.000  & 0.000  & 0.000  & 0.000  & 0.000  & 0.000  & 0.000   \\
			~ & 0.25 & 0.000  & 0.000  & 0.000  & 0.000  & 0.000  & 0.000  & 0.000  \\ \hline
		\end{tabular}
		\label{T41}	}
\end{table}

\begin{table}[H]
	\renewcommand\arraystretch{0.55}
	\caption{Empirical sizes and powers for $H_{42}$ in Study 4.}
	\center{
		\begin{tabular}{ccccccccc}
			\hline
			~ & \multirow{2}*{a} & n=100 & n=100 & n=100& n=100 & n=200 & n=400 & n=600 \\
			~ & ~ & p=2 & p=4 & p=6 & p=8 & p=12 & p=17 & p=20 \\ \hline
			~ & ~ & ~ & ~ & $q=p^2$ & ~ & ~ & ~ & ~ \\ \hline
			$Cost_n,\bSigma_1$ & 0.00 & 0.052  & 0.034  & 0.055  & 0.051  & 0.064  & 0.066  & 0.049 \\
			~ & 0.25 &0.324  & 0.381  & 0.401  & 0.416  & 0.755  & 0.960  & 0.999 \\ \hline
			$AICM_n,\bSigma_1$ & 0.00 & 0.038  & 0.000  & 0.000  & 0.000  & 0.000  & 0.000  & 0.000   \\
			~ & 0.25 & 0.329  & 0.000  & 0.000  & 0.000  & 0.000  & 0.000  & 0.000  \\ \hline
			\hline
			$Cost_n,\bSigma_2$ & 0.00 & 0.040  & 0.053  & 0.051  & 0.053  & 0.054  & 0.055  & 0.044  \\
			~ & 0.25 & 0.335  & 0.652  & 0.834  & 0.876  & 0.999  & 1.000  & 1.000   \\ \hline
			$AICM_n,\bSigma_2$ & 0.00 & 0.049  & 0.000  & 0.000  & 0.000  & 0.000  & 0.000  & 0.000  \\
			~ & 0.25 & 0.320  & 0.000  & 0.000  & 0.000  & 0.000  & 0.000  & 0.000  \\ \hline
			~ & ~ & ~ & ~ & $q=n$ & ~ & ~ & ~ & ~\\ \hline
			$Cost_n,\bSigma_1$ & 0.00 & 0.046  & 0.056  & 0.048  & 0.067  & 0.066  & 0.056  & 0.048 \\
			~ & 0.25 & 0.191  & 0.145  & 0.119  & 0.112  & 0.138  & 0.171  & 0.197 \\ \hline
			$AICM_n,\bSigma_1$ & 0.00 & 0.000  & 0.000  & 0.000  & 0.000  & 0.000  & 0.000  & 0.000   \\
			~ & 0.25 &0.000  & 0.000  & 0.000  & 0.000  & 0.000  & 0.000  & 0.000  \\ \hline
			\hline
			$Cost_n,\bSigma_2$ & 0.00 & 0.055  & 0.060  & 0.052  & 0.053  & 0.066  & 0.051  & 0.056  \\
			~ & 0.25 & 0.214  & 0.258  & 0.281  & 0.275  & 0.452  & 0.671  & 0.821  \\ \hline
			$AICM_n,\bSigma_2$ & 0.00 & 0.000  & 0.000  & 0.000  & 0.000  & 0.000  & 0.000  & 0.000   \\
			~ & 0.25 & 0.000  & 0.000  & 0.000  & 0.000  & 0.000  & 0.000  & 0.000 \\ \hline
		\end{tabular}
		\label{T42}	}
\end{table}

	\subsection{A real data example}
\begin{figure}[H]
		\centering
		\includegraphics[scale=0.5]{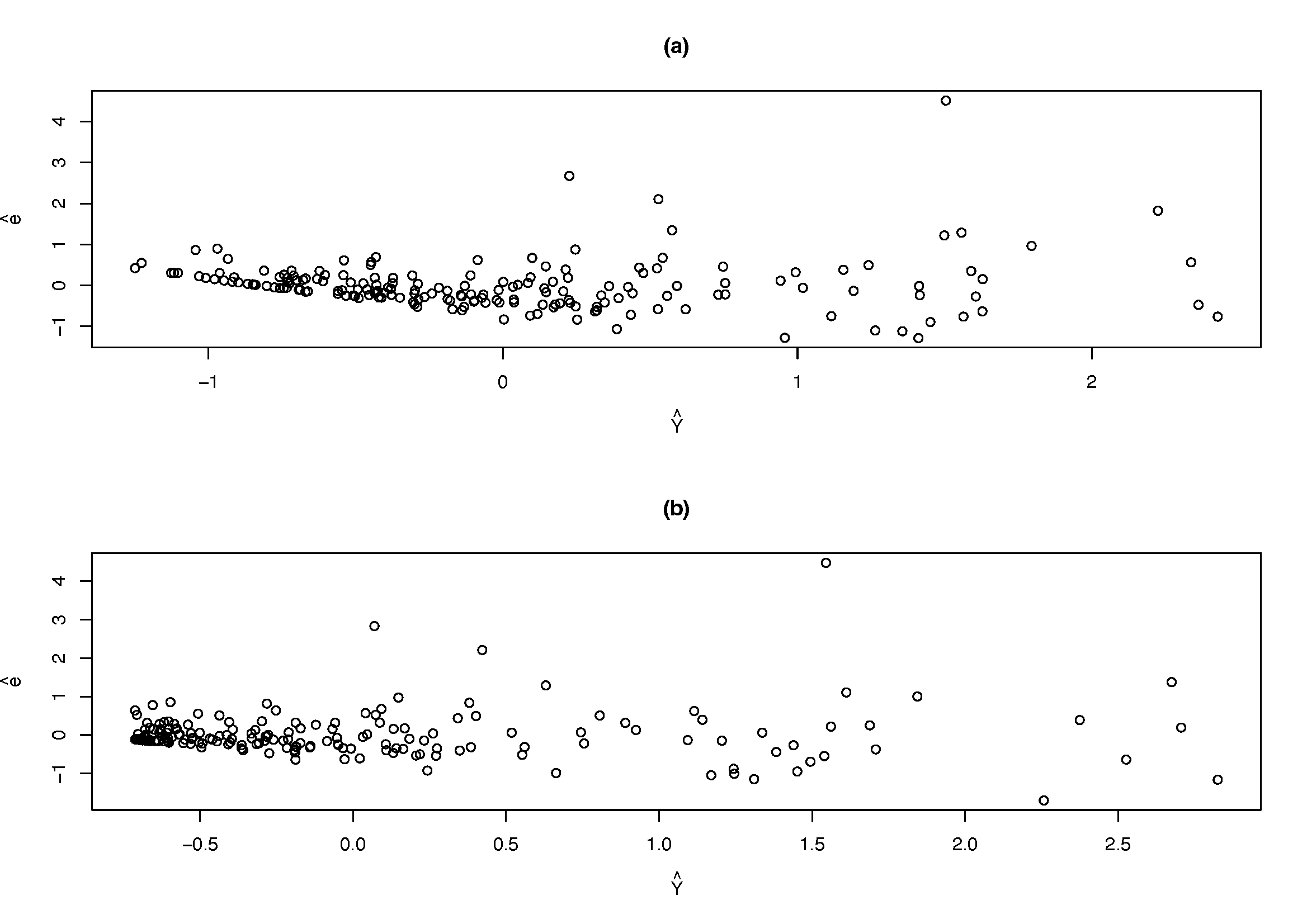}
		\caption{(a) Scatter plot of residuals generated from the linear regression model versus the fitted values,
			(b) Scatter plot of residuals generated from the model (\ref{e6})  versus the fitted values.}
		\label{P2}
	\end{figure}

	In this subsection, we use the CSM data set to illustrate our method. The CSM data set was first
	analyzed by \cite{ahmed2015using} and can be obtained through the website (\url{https://archive.ics.uci.edu/ml\\/datasets/CSM+\%28Conventional+and+
		Social+Media+Movies\%29+Dataset+2014+and\\+2015}). There are 187 observations left in the data set after cleaning 30 observations with missing responses and/or covariates.  The response variable Y is Gross Income. There are 11 predictor variables: Rating $X_1 $, Genre $X_2 $, Budget $X_3 $, Screens $X_4 $, Sequel $X_5 $, Sentiment $X_6 $, Views $X_7 $, Likes $X_8 $, Dislikes
	$X_9 $, Comments $X_{10} $ and Aggregate Followers $X_{11} $. To explore the relationship between the response Y and the predictor vector $ \boldsymbol{X}=(X_1,X_2,\dots,X_{11})^{\top}$, we first check whether the data set follows a linear regression model that is often used in practice. Figure \ref{P2} (a) shows that the residual plot might have a linear pattern. Besides, the value of our proposed test statistic $Cost_n $= 2.0420 and the $p$-value is about 0.0412. Therefore,  a linear regression model may not be tenable to fit the data, while  a more plausible model is in need. Using sufficient dimension reduction techniques like the cumulative slicing estimation (CSE), we find that the estimated structural dimension of this data set is $\hat{q}= 1,$ and the corresponding projected direction is
	$$\hat{\beta}=(0.3186, 0.0748, 0.5351, 0.5384, 0.0733, -0.0548, -0.2498, 0.2082, 0.3812, -0.0534, 0.2332)^{\top}.$$ As a result,  we establish the following polynomial regression model
	\begin{equation}
		\label{e6}
		Y=\theta_1+\theta_2\left(\bbeta^{\top}\boldsymbol{X}\right)+\theta_3\left(\bbeta^{\top}\boldsymbol{X}\right)^2+\varepsilon.
	\end{equation}
	The value of the test statistic $Cost_n$ is 0.7150, with the $p$-value being  0.4746,  indicating that the model (\ref{e6}) may be more appropriate to fit the CSM data set.  We also plot the residuals against the fitted responses in Figure \ref{P2} (b), which seems not to have a clear nonlinear pattern of the residuals. As this model is of a dimension reduction structure,  \cite{tan2022integrated}'s test also suggested this modeling.
	
	\section{Discussions}\label{sec5}
	This paper develops a novel test statistic for checking general parametric regression models in high-dimensional scenarios. By using a
	sample-splitting strategy and a conditional studentization approach,  the proposed test can obtain a normal limiting null distribution . It does not
	depend on dimension reduction model structures that are critically useful for existing tests. Moreover, our method is easy to implement, and does not need a resampling approximation for the critical value determination. The simulation results also show that
	our test, in many cases, can maintain the significance level and has good power performance. Thus, this
	research could be a good input in this research field. Further, it could be applied to other model–checking
	problems as a generic methodology.
	\par The sample-splitting technique also brings some limitations. The main shortcoming is that the test statistic converges to its weak limit at the rate of $1/\sqrt {n_1}$ rather than $1/\sqrt {n}$ causing some loss in power. Thus, the sample size should not be too small otherwise this methodology may not work well.    We note that the commonly used cross-fitting idea is often useful in other testing problems to enhance power.   But the following observations make us hesitant to use this method. Although the  studentisation approach considers the conditional variance in denominator when the second subset of data is given, the test involves all datum points even in the conditional variance estimation. If we construct  another conditional studentized test with given the first subset of data,  the  numerators of the two test statistics are the same, but  the denominators  are highly correlated and the covariance is hard to compute, if not impossible.  As the limiting null distribution is intractable in this case, a possible solution is to use re-sampling approximation such as the wild bootstrap. We tried this idea for Model 1 in Study~1, and found that the power can be enhanced, but slightly.  But such a solution  gives up  the main advantage of our method having the tractable limiting null distribution. It may not be as valuable as that.
	Another issue is about the model dimensions. In our setting, without  sparsity structure, the method can handle the cases where the number of the parameters can diverge at the rate of order $n^{1/3}$ and most likely this rate cannot be higher (see, e.g., \cite{tan2022integrated}). To handle the cases with larger number of parameters, model sparsity assumption about parameters could be necessary. See the relevant references \cite{shah2018goodness} and \cite{jankova2020goodness} that checked  sparse linear and generalized linear models. But in these cases, the construction of the test statistic may have to use penalization method for variable selection,  thus in the more general model setting than those in \cite{shah2018goodness} and \cite{jankova2020goodness}, it is still unclear whether the asymptotic behaviors can be derived and even if we can, it is still being determined whether  the asymptotic distribution-free property holds. These deserve further study. On the other hand, it is of interest to see that when some conditions on  regression function holds, the dimension of the predictor vector can be high, even higher than  the sample size, although the conditions are stringent in the large $q$ scenarios.  The simulations with $q=p^2$ and $q=n$  support this observation. We also conducted some simulations with $q$ being larger than $n$, the phenomenon is similar. This shows the potential of our method tackling the models with higher dimensions of the predictor vector.
	
	\section{Regularity Conditions}\label{sec6}
	In these conditions, $\|\cdot\|$ means the $L_2$ norm of any vector or matrix.
	\begin{con}
		\label{CA.1}
		There exists a unique minimizer $\boldsymbol{\theta}^*\in \mathbb{R}^p$  of the square loss in the interior of the compact parameter set $\boldsymbol{\Theta}$.
	\end{con}
	
	\begin{con}
		\label{CA.2} Denote $\boldsymbol{\theta}=\left(\theta_1,\cdots,\theta_p\right)$.
		The regression function admits third derivatives concerning $\boldsymbol{\theta}\in\boldsymbol{\Theta}$. For $\forall j,k=1,2,\cdots,p$, define
		$$\dot{g}(\boldsymbol{\theta},\boldsymbol{x})=\frac{\partial g(\boldsymbol{\theta},\boldsymbol{x})}{\partial\boldsymbol{\theta}}, \ddot{g}(\boldsymbol{\theta},\boldsymbol{x})=\frac{\partial^2 g(\boldsymbol{\theta},\boldsymbol{x})}{\partial\boldsymbol{\theta}\partial\boldsymbol{\theta}^{\top}}, \dot{g}_j(\boldsymbol{\theta},\boldsymbol{x})=\frac{\partial g(\boldsymbol{\theta},\boldsymbol{x})}{\partial\theta_j}  \text{ and }   \ddot{g}_{jk}(\boldsymbol{\theta},\boldsymbol{x})=\frac{\partial g(\boldsymbol{\theta},\boldsymbol{x})}{\partial\theta_j\partial\theta_k}.$$
		Let $U(\boldsymbol{\theta}^*)$ be a subset  consists of all $\boldsymbol{\theta}$ satisfying $\|\boldsymbol{\theta}-\boldsymbol{\theta}^*\|\leq r_0$ in the interior of $\boldsymbol{\Theta}$, where $r_0$ is a positive constant. Further, for $\forall j,k=1,2,\cdots,p$ and $\forall\boldsymbol{\theta}\in U(\boldsymbol{\theta}^*)$, there exists a function $F(\cdot)$ such that for any $\boldsymbol{x}$,
		$|g(\boldsymbol{\theta},\boldsymbol{x})|\leq F(\boldsymbol{x}), |\dot{g}_{j}(\boldsymbol{\theta},\boldsymbol{x})|\leq F(\boldsymbol{x}), |\ddot{g}_{jk}(\boldsymbol{\theta},\boldsymbol{x})|\leq F(\boldsymbol{x})$
		with $E\left\{F^4(\boldsymbol{X})\right\}=O\left(1\right)$. The fourth moment  $E(\varepsilon^4)$ of $\varepsilon$ is bounded for the regression model.
	\end{con}
	\begin{con}\label{CA.31}
		For $j=1,2,\cdots,p$, define
		$\psi_{j}(\boldsymbol{\theta},\boldsymbol{x})=\left\{m(\boldsymbol{x})-g(\boldsymbol{\theta},\boldsymbol{x})\right\}\dot{g}_{j}(\boldsymbol{\theta},\boldsymbol{x})$,
		and	$P\ddot{\psi}_{j\boldsymbol{\theta}}=E\left\{\frac{\partial\psi_{j}}{\partial \boldsymbol{\theta}\partial \boldsymbol{\theta}^{\top}}(\boldsymbol{\theta},\boldsymbol{X})\right\}$. Let
		$\lambda_i\left(P\ddot{\psi}_{j\boldsymbol{\theta}}\right)$ be the $i$-th eigenvalue of $P\ddot{\psi}_{j\boldsymbol{\theta}}$.
		$\max _{1 \leq i, j \leq p}\lambda_i\left(P\ddot{\psi}_{j\boldsymbol{\theta}}\right)\leq c$, for any $\boldsymbol{\theta}\in U(\boldsymbol{\theta}^*),$
		where  $c$ is a positive constant free from $n$ and $p$.
	\end{con}
	
	For any matrix $\boldsymbol{A}$, let $\lambda_{\min}(\boldsymbol{A})$ and $\lambda_{\max}(\boldsymbol{A})$ be the smallest and the largest eigenvalue of $\boldsymbol{A}$, respectively.
	\begin{con}
		\label{CA.3}
		
		Define $\bSigma=E\left\{\dot{g}(\btheta^*, \boldsymbol{X}) \dot{g}(\btheta^*, \boldsymbol{X})^{\top}\right\}-E\left[\left\{m(\boldsymbol{X})-g(\btheta^*,\boldsymbol{X})\right\} \ddot{g}(\btheta^*, \boldsymbol{X})\right]$. 		
		There exist two constants $a$ and $b$ unrelated to $n$ and $p$, such that
		$0<a\leq\lambda_{\min}(\bSigma)\leq \lambda_{\max}(\bSigma)\leq b<\infty.$
		
		Besides, define $\bSigma_*=E\left\{\dot{g}(\boldsymbol{\theta}^*,\boldsymbol{X})\dot{g}(\boldsymbol{\theta}^*,\boldsymbol{X})^{\top}\right\}$ 	and $\bSigma_{\varepsilon}=E\left\{\varepsilon^2\dot{g}(\boldsymbol{\theta}^*,\boldsymbol{X})\dot{g}(\boldsymbol{\theta}^*,\boldsymbol{X})^{\top}\right\}$, where $\varepsilon$ is the error term corresponding to $\boldsymbol{X}$ in our regression model. $a^*$, $b^*$, $a_{\varepsilon}$ and $b_{\varepsilon}$ are four constants unrelated to $n$ and $p$ such that
		$0<a^*\leq\lambda_{\min}(\bSigma_*)\leq \lambda_{\max}(\bSigma_*)\leq b^*<\infty,$
		and
		$0<a_{\varepsilon}\leq\lambda_{\min}(\bSigma_{\varepsilon})\leq \lambda_{\max}\left(\bSigma_{\varepsilon}\right)\leq b_{\varepsilon}<\infty.$
		
		Finally,	define $\bSigma^*_{cov}=E\left\{\dot{g}(\boldsymbol{\theta}^*,\boldsymbol{X}_1)\dot{g}(\boldsymbol{\theta}^*,\boldsymbol{X}_2)^{\top}w_{12}\right\}$ and assume $\|\bSigma^*_{cov}\|\leq b_{cov}$, where $b_{cov}$ is a constant free from $n$ and $p$.
	\end{con}
	\begin{con}
		\label{CA.4}
		Define
		$\ddot{\psi}_{jkl}(\boldsymbol{\theta},\boldsymbol{x})=\frac{\partial^{2} \psi_{j}}{\partial \theta_{k} \partial \theta_{l}}(\boldsymbol{\theta},\boldsymbol{x})\   \text{ and }\  {g}_{jkl}^{(3)}(\boldsymbol{\theta},\boldsymbol{x})=\frac{\partial^{3} g}{\partial\theta_{j} \theta_{k} \partial \theta_{l}}(\boldsymbol{\theta},\boldsymbol{x}).$
		There exist two measurable functions $F_1(\cdot)$ and $F_2(\cdot)$ with $E\left\{F_1^4(\boldsymbol{X})\right\}=O\left(1\right)$ and $E\left\{F_2^8(\boldsymbol{X})\right\}=O\left(1\right)$, satisfying
		$
		\left|\ddot{\psi}_{jkl}(\boldsymbol{\theta}_1,\boldsymbol{x})-\ddot{\psi}_{jkl}(\boldsymbol{\theta}_2,\boldsymbol{x})\right| \leq \sqrt{p}\|\boldsymbol{\theta}_{1}-\boldsymbol{\theta}_{2}\| F_1(\boldsymbol{x})
		$ and
		$
		\left|g_{jkl}^{(3)}(\boldsymbol{\theta}_{1}, \boldsymbol{x})-g_{jkl}^{(3)}(\boldsymbol{\theta}_{2}, \boldsymbol{x})\right| \leq \sqrt{p}\|\boldsymbol{\theta}_{1}-\boldsymbol{\theta}_{2}\| F_2(\boldsymbol{x}),
		$
		where $\boldsymbol{\theta}_1,\boldsymbol{\theta}_2\in U(\boldsymbol{\theta}^*)$.
	\end{con}
	\begin{con}\label{lip}	There exists a measurable function $L(\cdot)$ with $E\left\{L^4(\boldsymbol{X})\right\}=O\left(1\right)$,
		satisfying
		\begin{align*}
			\left\|\dot{g}(\boldsymbol{\theta}_1,\boldsymbol{x)}-\dot{g}(\boldsymbol{\theta}_2,\boldsymbol{x})\right\|\leq \sqrt{p}\|\boldsymbol{\theta}_1-\boldsymbol{\theta}_2\|L(\boldsymbol{x}),
		\end{align*}
		where $\boldsymbol{\theta}_1, \boldsymbol{\theta}_2\in U(\boldsymbol{\theta}^*)$.
	\end{con}
	\begin{remark}
		\label{ReA.2}
		Conditions \ref{CA.1}, \ref{CA.2}, \ref{CA.31} and \ref{CA.4} appear in \cite{tan2022integrated}, commonly assumed in high dimensional model checking literature. Condition \ref{CA.3} is similar to the regularity condition (A2) in \cite{tan2022integrated}, while we also assume that the largest eigenvalue and the smallest eigenvalue are bounded in the other three matrices. Condition \ref{lip} is a general Lipschitz Condition.
		For Conditions \ref{CA.1}-\ref{lip}, we do not directly put the constraints on the dimension $q$ of the predictor vector $\boldsymbol{X}$, while they are hidden in the boundedness of the related functions and their derivatives. Though our conditions could be stringent when $q$ diverges quickly to infinity, some functions still meet the requirements. Thus, the test could work with a high-dimensional predictor vector.
	\end{remark}
	Recall that $W_n(\boldsymbol{x},\boldsymbol{z})$ in our paper is a function concerning $\boldsymbol{x}$ and $\boldsymbol{z}$, and without confusion and mis-understanding,  $w_{ij}=W_n(\boldsymbol{X}_i,\boldsymbol{X}_j)$ for the sake of simplicity.
	\begin{con}\label{w}
		$\inf_{\boldsymbol{x}_i,\boldsymbol{x}_j}W_n(\boldsymbol{x}_i,\boldsymbol{x}_j)\geq 0$ and $ E\left\{W_n^4(\boldsymbol{X}_1,\boldsymbol{X}_2)\right\}=E\left(w_{12}^4\right)\leq C_1$ for a positive constant $C_1$. 
	\end{con}
	\begin{remark}\label{remark2}
		For Condition \ref{w},
		$w_{12}$, as a weight function of $\boldsymbol{X}_1$ and $\boldsymbol{X}_2$,  can satisfy $E\left(w_{12}^4\right)\leq C_1$ in many forms, such as $w_{12}=1/\sqrt{\|\boldsymbol{X}_1-\boldsymbol{X}_2\|^2/q+1}$ and $w_{12}=\exp(-\|\boldsymbol{X}_1-\boldsymbol{X}_2\|^2/2q)$. In both two cases, $w_{12}\leq 1$ always holds. For the form like $w_{12}=\sum_{k=1}^{q}\exp\left\{-(X_{1k}-X_{2k})^2/2\right\}$, although $E\left(w_{12}^4\right)$ may diverge to infinity as $q\to\infty$,  we can divide both numerator and denominator by $q$ and use $w^{*}_{12}=w_{12}/q$ to replace $w_{12}$, then $E\left({w^{*}_{12}}^4\right)\leq 1$.
		In fact, under Conditions \ref{CA.3} and \ref{w}, we can infer that $\|E\left\{\dot{g}(\boldsymbol{\theta}^*,\boldsymbol{X}_1)\right\}\|$ and  $\|E_{\boldsymbol{X}_1}\left\{\dot{g}(\boldsymbol{\theta}^*,\boldsymbol{X}_2)w_{12}\right\}\|$ are bounded. In  Supplementary Material, we give some more details to show this condition is satisfied.
	\end{remark}
	\begin{con}\label{bc}
		$\max_{1\leq k,l\leq p}E\left\{\varepsilon_1^4\dot{g}_k^2(\boldsymbol{\theta}_0,\boldsymbol{X}_1)\dot{g}_l^2(\boldsymbol{\theta}_0,\boldsymbol{X}_1)\right\}<\infty.$
	\end{con}

	Define
	\begin{align}\label{w111} w{'}_{ij}=&w_{ij}-\dot{g}(\boldsymbol{\theta}^*,\boldsymbol{X}_j)^{\top}\bSigma^{-1}E_{\boldsymbol{X}_i}\left\{\dot{g}(\boldsymbol{\theta}^*,\boldsymbol{X}_j)w_{ij}\right\}-\dot{g}(\boldsymbol{\theta}^*, \boldsymbol{X}_{i})^{\top}\bSigma^{-1}E_{\boldsymbol{X}_j}\left\{\dot{g}(\boldsymbol{\theta}^*,\boldsymbol{X}_i)w_{ij}\right\}\nonumber\\&+\dot{g}(\boldsymbol{\theta}^*, \boldsymbol{X}_{i})^{\top}\bSigma^{-1}\bSigma_{cov}^*\bSigma^{-1}\dot{g}(\boldsymbol{\theta}^*,\boldsymbol{X}_j).
	\end{align}
	
	\begin{con}\label{r1}
		$E\left(\varepsilon_1^2\varepsilon_2^2w{'}_{12}^2\right)\geq C_3$, and $E\left(\varepsilon_1^4\varepsilon_2^4w{'}_{12}^4\right)\leq C_4$, where $C_3$ and $C_4$ are two positive constants. Besides,
		$E\left[\left\{\dot{g}(\boldsymbol{\theta}_0, \boldsymbol{X}_1)^{\top}\bSigma^{-1}\bSigma_{cov}^*\bSigma^{-1}\dot{g}(\boldsymbol{\theta}_0,\boldsymbol{X}_2)\right\}^4\right]=O\left(1\right).
		$
	\end{con}
	\begin{remark} Condition \ref{r1} is a sufficient condition for the Berry-Esseen bound, which is essential to ensure the asymptotic normality of our test statistic under the null hypothesis.
		For Condition \ref{r1}, let's analyse the meaning of $w{'}_{ij}$. The linear projection of $w_{ij}$ on $\dot{g}(\boldsymbol{\theta}^*,\boldsymbol{X}_i)$ is $\dot{g}(\boldsymbol{\theta}^*, \boldsymbol{X}_{i})^{\top}\bSigma^{-1}E_{\boldsymbol{X}_j}\left\{\dot{g}(\boldsymbol{\theta}^*,\boldsymbol{X}_i)w_{ij}\right\}$, and the linear projection of the remainder term,  $w_{ij}-\dot{g}(\boldsymbol{\theta}^*, \boldsymbol{X}_{i})^{\top}\bSigma^{-1}E_{\boldsymbol{X}_j}\left\{\dot{g}(\boldsymbol{\theta}^*,\boldsymbol{X}_i)w_{ij}\right\}$, on $\dot{g}(\boldsymbol{\theta}^*,\boldsymbol{X}_i)$, is $\dot{g}(\boldsymbol{\theta}^*,\boldsymbol{X}_i)\bSigma^{-1}E_{\boldsymbol{X}_i}\left\{\dot{g}(\boldsymbol{\theta}^*,\boldsymbol{X}_j)w_{ij}\right\}-\\ \dot{g}(\boldsymbol{\theta}^*,\boldsymbol{X}_i)\bSigma^{-1}E\left\{\dot{g}(\boldsymbol{\theta}^*,\boldsymbol{X}_1)\dot{g}(\boldsymbol{\theta}^*,\boldsymbol{X}_2)^{\top}w_{12}\right\}\bSigma^{-1}\dot{g}(\boldsymbol{\theta}^*,\boldsymbol{X}_j)$. Then we can find that $w{'}_{ij}$ has the form of the remainder term after $w_{ij}$  projected on $\dot{g}(\boldsymbol{\theta}^*,\boldsymbol{X}_i)$ and $\dot{g}(\boldsymbol{\theta}^*,\boldsymbol{X}_j)$.
		Hence $w{'}_{ij}$ is almost the estimation of the error term when we use $\dot{g}(\boldsymbol{\theta}^*,\boldsymbol{X}_i)$ and $\dot{g}(\boldsymbol{\theta}^*,\boldsymbol{X}_j)$ to predict $w_{ij}$ using a linear model. Since the structure of $w_{ij}$ and $\dot{g}(\boldsymbol{\theta}^*,\boldsymbol{X}_i)$ are different, it is reasonable to assume that $E\left(\varepsilon_1^2\varepsilon_2^2w{'}_{12}^2\right)$ has a lower bound as $p$ goes to infinity. Since $w_{ij}$ is bounded, it is reasonable  to assume the boundedness of $E\left(\varepsilon_1^4\varepsilon_2^4w{'}_{12}^4\right)$.
	\end{remark}
	Define $\bSigma_l=E
	\left\{l^2(\boldsymbol{X})\dot{g}\left(\boldsymbol{\theta}_0, \boldsymbol{X}\right)\dot{g}\left(\boldsymbol{\theta}_0, \boldsymbol{X}\right)^{\top}\right\}$. Under the global hypothesis, $l(\boldsymbol{X})=m(\boldsymbol{X})-g(\boldsymbol{\theta}^*,\boldsymbol{X})$.
	\begin{con}\label{r2} $E\left\{l^4(\boldsymbol{X})\right\}=O\left(1\right)$.
		Further, there are two constants $a_l$, $b_l$ not depending on $n$ and $p$ such that
$0<a_l\leq \lambda_{\min}(\bSigma_l)\leq \lambda_{\max}(\bSigma_l)\leq b_l<\infty.$
	\end{con}
	
	
	Recall the sequence of local alternatives, $H_{1 n}: Y=g(\boldsymbol{\theta}_0,\boldsymbol{X})+\delta_{n} l(\boldsymbol{X})+\varepsilon$,
	and the definition of $H(\boldsymbol{X}_1)$ in Subsection~\ref{sec3.2}.
	
	\begin{con}\label{r3}
		$H(\boldsymbol{X}_1)\neq 0, a.s.$ , $E\left\{|\varepsilon_1H(\boldsymbol{X}_1)|^2\right\}\geq C_5> 0$, a.s., and $E\left\{|\varepsilon_1H(\boldsymbol{X}_1)|^4\right\}\leq C_6<\infty$, where $C_5$ and $C_6$ are two constants unrelated to $n$ and $p$.
	\end{con}
	\begin{remark}Conditions \ref{r2} and \ref{r3} are used to obtain the asymptotic distribution of our test statistic under the alternative hypotheses.
		These two conditions are not necessary for studying the test statistic with high power under the alternative hypotheses. Conversely, if we
		do not put conditions either on the moments of $l(\boldsymbol{X})$ and $H(\boldsymbol{X})$ or on an upper bound of the eigenvalues of  $\bSigma_l$, our test statistic may diverge at a faster rate to have higher power. On the other hand, when we want to study the properties under the local alternatives,   these conditions can make the investigation of the limiting properties of the test statistic more easily with some easily understood presentations. 
		
	\end{remark}

	\begin{center}
		{\large\bf SUPPLEMENTARY MATERIAL}
	\end{center}

	\begin{description}
		\item {\bf Supplementary of  The conditionally studentized test for high-dimensional parametric regressions} Technical proofs of the theorems. (.pdf file)
		
	\end{description}
	
	\bibliographystyle{abbrvnat}
	\bibliography{reference}
	
\end{document}